\documentclass[10pt]{article}
\usepackage[english]{babel}
\usepackage[utf8x]{inputenc}
\usepackage[OT1]{fontenc}
\usepackage{amsfonts, amsmath, amsthm, amssymb}
\usepackage{graphicx}
\usepackage{listings}

\usepackage{multicol} 
\usepackage{amssymb, amsmath, amsbsy}
\usepackage{stackrel}

\usepackage{subcaption}
\usepackage{graphicx}

\usepackage{listings}
\usepackage{csquotes}
\usepackage[hidelinks]{hyperref}
\usepackage[nottoc,notlot,notlof]{tocbibind}

\usepackage[margin=0.8in]{geometry}
\usepackage{xcolor}
\usepackage{braket}

\title{Corrected Hill Function in Stochastic Gene Regulatory Networks}

\author{Manuel Eduardo Hernández-García, Jorge Velázquez-Castro \\
       Facultad de Ciencias Físico Matemáticas \\
       Benemérita Universidad Autónoma de Puebla}

\date{July 2023}

\begin{document}
\maketitle 

\abstract{Describing reaction rates in stochastic bio-circuits is commonly done by directly introducing the deterministically deduced Hill function into the master equation. However, when fluctuations in ligand-receptor reaction rates are not negligible,  the Hill function must be derived, considering all stochastic reactions involved. In this study, we derived the stochastic version of the Hill function from the master equation of the complete set of reactions that, in the macroscopic limit, lead to the Hill function reaction rate. We performed a series expansion around the average values of the concentrations, which allowed us to determine corrections for the deterministic Hill function. This methodology enabled us to quantify the fluctuations associated with ligand-receptor reactions. We found that the underlying variability in the propensity rates of gene regulatory networks has an important nonlinear effect that reduces the intrinsic fluctuations of mRNA and protein concentrations.


}

\tableofcontents

\newpage

\section{Introduction}\label{sec1}
The study of genetic regulatory networks has increased in recent years, owing to their potential applications in novel disease treatments. Understanding the dynamics and functional relationships between the components of the regulatory networks of some genetic diseases \cite{Buloa, Alon} can provide insights into their causes and lead to new treatments.  These systems are complex and contain several components. However, the transcription factors and protein concentrations involved in network dynamics are generally low. The effect of fluctuations around average concentrations can propagate between many components or even play a functional role.  Thus, a deterministic description of biochemical reactions is an approximation. Many models have been proposed in the context of stochastic processes \cite{Wal, Paul} using the master equation. Most of these approaches employ a deterministically derived Hill function to represent certain chemical rates, implying that they are approximations of a complete reaction network. This problem was previously addressed by \cite{Thomas, Hole}, and in \cite{Thomas, Uri, Santi, Kim} important progress in this area has been made; however, its consequences and the problem itself are yet to be exhaustively studied. Although in \cite{Lip} a Hill function was derived in a stochastic environment that was only applied to specific conditions and systems, the generalization of the stochastic Hill function derived in this study can be applied to a wider range of cases. 

The formalism of multivariable birth-dead processes \cite{Gar} is typically employed to describe reaction kinetics. However, in addition to the simple reactions, the master equation is difficult to solve analytically. Furthermore, as the reaction network becomes more complex, numerical solutions become more computationally intensive.

 This lack of practicality in finding a solution to the master equation results in a linear noise approximation that leads to the Fokker-Planck equation, which is the standard framework for making inferences about stochastic systems \cite{Gar, Tom}. In \cite{Gomez}, a different method was explored, in which only second-order reactions were investigated. In our study, we utilized a comparable approach but extended it to cover models of any order. We accomplish this by examining sufficiently large systems, where the resulting dynamics feature a corrective term. To determine the Hill function using this approximation, we leveraged the distinction between rapid and slow reactions in certain situations \cite{Uri, Gout, Kim2, Cao}, as some biochemical processes, such as enzymatic reactions, occur more rapidly.
  
 In this study, we obtained general expressions for the master equation based on the assumption that fast and slow reactions exist in the chemical network, where the fast reactions have already reached their equilibrium distributions. This procedure provides a more accurate description than inserting the Hill function to describe the ligand-receptor response, which is commonly performed.

The master equation for slow reactions was then used to obtain the evolution of the average reactant concentrations as a series of expansions of relative fluctuations.  Because mean concentrations are associated with deterministic reaction kinetics, this procedure allows us to correct the deterministic dynamics of small systems in which the impact of intrinsic fluctuations cannot be disregarded.

 This method was illustrated using three examples. First, we found the corrections of the Hill function due to reactant concentration fluctuations in small systems using a Toggle Switch. With this gained experience, an expression for the master equation of a general gene regulatory network was provided and used to analyze the repressilator and the activator-repressor clock. 
  
The proposed methodology helps find a more accurate description of complex reaction kinetics for small systems than deterministic models. This is also an improvement over the simplistic application of the Hill function for describing the reaction rates of ligand-receptor reactions in a stochastic manner. Additionally, the resulting ordinary differential equations are significantly more computationally efficient than expensive computer simulations of the Gillespie algorithm \cite{Gillespie, Lecca}. This approximation will allow for simulation and thus, a better understanding of more complex systems and the role of fluctuations in chemical networks. In particular, it is possible to analyze the effect of intrinsic noise in complex gene regulatory networks and the potential emergent properties from the inherent noise in these types of systems.\\

The remainder of this paper is organized as follows.

In Section  \ref{section2}, we briefly describe the multivariable birth-death processes and describe a method for obtaining deterministic equations from a stochastic model, particularly for multivariable birth-death systems. This approximation allowed us to quantify fluctuations in the variables throughout the time evolution. 

Section \ref{section4} briefly explains the emergence of the Hill function from a stochastic process. In particular, we first analyzed the toggle switch to explain in more detail how the Hill function was obtained. To achieve this, we assume both fast and slow reactions. Using a similar procedure, we calculated the Hill function with corrections owing to stochastic effects.

Section \ref{section5} presents a generic gene regulation network. This study is divided into two broad categories. First, we present a general network with one transcription factor, and then the study is broadened to consider many transcription factors. 
 
In Section \ref{section6}, we review the results and provide concluding remarks.


\section{Chemical master equation and the dissipation fluctuation theorem}\label{section2}
In deterministic models, the law of mass action is used to build a set of differential equations that describe the concentration dynamics of the species for a network of chemical processes. However, this approach may not always be the most suitable for small systems (low molecule numbers), because the relative fluctuations become significant. This is what we mean by "small" in this context, when the fluctuations are comparable with the mean itself. The mass action law can be generalized in the stochastic realm as a multivariable birth-dead process \cite{Gar}. In this section, the methodology and notation used to describe the birth-dead process are reviewed.

We consider $N$ species $\mathcal{S}_j$ ($j$ $\in$ {$1,2,..N$}), and $M$ reactions $\mathcal{R}_i$ ($i$ $\in$ {$1,2,..M$}) through which these species are transformed, that is,
{\small
\begin{align}
    \mathcal{R}_i : \sum_{j=1}^{N} \alpha_{ij} \mathcal{S}_j \stackbin[k_{i}^{-}]{k_{i}^{+}}{\rightleftarrows} \sum_{j=1}^{N} \beta_{ij} \mathcal{S}_j. \label{2.1}
\end{align}}
The coefficients $\alpha_{ij} $ and $ \beta_{ij}$ are positive integers known as stoichiometric coefficients. The stoichiometric matrix is defined as
\begin{align}
    \Gamma_{ji}= \beta_{ij} -  \alpha_{ij} .
\end{align}
Through collisions (or interactions) of the species, they may be transformed; therefore, the transformation rates are proportional to collision probability. The propensity rates can be expressed as follows  \cite{Gar}
{\small
\begin{align}
    {t_i^{+}(\mathbf{S})}&= k_{i}^{+} \prod _{l} \frac{S_l !}{\Omega ^{\alpha_{il}}(S_l- \alpha_{il} )!}, &
    {t_i^{-}(\mathbf{S})}= k_{i}^{-} \prod _{l} \frac{S_l !}{\Omega ^{\beta_{il}}(S_l- \beta_{il} )!}. 
\end{align}}
(Index $i$ denotes the reaction that occurs $\mathcal{R}_i$). Where $S_l$ is the number of molecules of species $\mathcal{S}_l$ and $\mathbf{S} = (S_1,S_2,..., S_N)$ is the state vector.  These ratios represent the transition probabilities between different states of the system. $\Omega$ is closely related to the system size. According to (\ref{2.1}), we label the reactions that go from left to right as $t^{+}_i(\mathbf{S})$ and when they move in the opposite direction as $t^{-}_i(\mathbf{S})$. With all of the above elements, we can write the master equation of the system, which describes the dynamic evolution of the probability distribution of the states of the system. The master equation can then be expressed as

{\small
\begin{align}
    \partial_{t} P(\mathbf{S},t)= \Omega \sum _i &\left( t_i^{-}(\mathbf{S}+\Gamma_{i}) P(\mathbf{S}+\Gamma_{i},t) - t_i^{+}(\mathbf{S}) P(\mathbf{S},t) \right. \nonumber \\
    &+\left.t_i^{+}(\mathbf{S}-\Gamma_{i}) P(\mathbf{S}-\Gamma_{i},t) - t_i^{-}(\mathbf{S}) P(\mathbf{S},t) \right).  \label{2.4}
\end{align}}
Where $\Gamma_{i}$ denotes the $i$-column of the stoichiometric matrix.  The sum is calculated for all the reactions of the system. This equation is also known as the master chemical equation and is the stochastic version of the law of mass action. 

A system with fluctuations can be directly analyzed using a master equation. However, one of our goals is to quantify fluctuations; therefore, it is more efficient to use an approximation than to directly use the master equation. For example, the approximation used is similar to that presented in \cite{Gomez}. In this case, to obtain deterministic equations, we used a second-order expansion around the average for sufficiently large $\Omega$ or first-order systems as follows: 

{\tiny
\begin{align}
    \braket{f(\mathbf{X})} \approx &  \left\langle f(\braket{\mathbf{X}}) + \sum_i (X_i-\braket{X_i})  \left[ \frac{\partial f(\mathbf{X})}{\partial  X_i} \right]_{\mathbf{X}=\braket{\mathbf{X}}}  +  \sum_i \sum_j \frac{(X_i-\braket{X_i})(X_j-\braket{X_j})}{2} \left[ \frac{\partial^2 f(\mathbf{X})}{\partial X_i \partial X_j} \right]_{\mathbf{X}=\braket{\mathbf{X}}}\right\rangle \nonumber \\
    = & f(\braket{\mathbf{X}})  +  \sum_i \sum_j  \frac{\sigma^2(X_i,X_j)}{2} \left[\frac{\partial^2 f(\mathbf{X})}{\partial X_i \partial X_j}\right]_{\mathbf{X}=\braket{\mathbf{X}}}, \label{3.13}
\end{align}}
where $\braket{\mathbf{X}} = (\braket{X_1},\braket{X_2},..., \braket{X_N})$ and $   \sigma^2(X_i,X_j)=\left\langle (X_i-\braket{X_i})(X_j-\braket{X_j})\right\rangle $.

Using the master equation of the system (\ref{2.4}) and the approximation of equation (\ref{3.13}), we obtain the evolution of concentration ($s_{j}$=$\frac{\braket{S_j}}{\Omega}$), 

{\footnotesize
\begin{align}
    \frac{d s_j }{d{t}}=& \sum_{i} \Gamma_{ji}\left( R_i^{D+}(\mathbf{s}) - R_i^{D-}(\mathbf{s}) + \sum_{j_1} \sum_{j_2}  \frac{\sigma^2(s_{j_1}, s_{j_2})}{2}  \frac{\partial^2}{\partial {s_{j_1}}\partial {s_{j_2}}} (R_i^{D+}(\mathbf{s}) - R_i^{D-}(\mathbf{s})) \right). \label{3.19}
\end{align}}
(we denote  $t_i^{+}(\braket{\mathbf{S}})\approx k_{i}^{+} \prod _{j} s_j^{\alpha_{ij}}=R_i^{D+} ({\mathbf{s}})$ and $t_i^{-}(\braket{\mathbf{S}})\approx k_{i}^{-} \prod _{j} s_j^{\beta_{ij}}=R_i^{D-} ({\mathbf{s}})$ as the deterministic reaction rates) This equation provides the first stochastic correction to the deterministic evolution of the reaction kinetics. However, this is not yet a closed system of equations; we also need the evolution of the covariance $\sigma^2(s_{j_1}, s_{j_2})$ to solve it.  As before, we use the master equation and then we use the expansion (\ref{3.13}) to obtain 

{\footnotesize
\begin{align}
    \frac{\partial}{\partial{t}} \sigma^2(s_{l_1}, s_{l_2})&=    \sum _i  \left(  \Gamma_{l_1 i}\Gamma_{l_2i}\frac{(R_i^{D+}(\mathbf{s})+R_i^{D-}(\mathbf{s}))}{\Omega} \right. \nonumber \\
    &  + \sum_{j_1}   \left( {\sigma^2(s_{l_1}, s_{j_1})} \Gamma_{l_2 i} \frac{\partial}{\partial {s_{j_1}}} +{\sigma^2(s_{j_1}, s_{l_2})} \Gamma_{l_1 i} \frac{\partial}{\partial{s_{j_1}}}\right) {(R_i^{D+}(\mathbf{s})-R_i^{D-}(\mathbf{s}))}  \nonumber \\
    &  + \left. \sum_{j_1} \sum_{j_2}  \left(\sigma^2(s_{j_1},  s_{j_2}) \frac{\Gamma_{l_1 i}\Gamma_{l_2i}}{2 \Omega} \frac{\partial^2}{\partial {s_{j_1}}\partial {s_{j_2}}} (R_i^{D+}(\mathbf{s})+R_i^{D-}(\mathbf{s}))  \right) \right), \label{3.24}
\end{align}}
It is worth noting that in contrast to the common Linear Noise Approximation, the cross-reaction terms from Equations (\ref{3.19}) and (\ref{3.24}) provide a more accurate description of the system \cite{Ramon}.

As the intrinsic fluctuations of the species concentrations are given by
{\small
\begin{align}
    \eta^2_i = \frac{\braket{\xi_i^{2}}}{\Omega^2}= \frac{\sigma^2(X_i,X_i)}{\Omega^2} , \label{3.16}
\end{align}}
where $\xi_i^{2}= (X_i-\braket{X_i})^2$.
Subsequently, the system of differential equations formed by (\ref{3.19}) and (\ref{3.24}) describe the mean dynamics of the system and quantify its intrinsic fluctuations. Thus, instead of solving the entire master equation, it is possible to solve the more simple equations (\ref{3.19}) and (\ref{3.24}). We will refer to this set of differential equations as MFK (Mass Fluctuation Kinetics). In the following sections, we describe some representative stochastic systems and highlight the advantages of this description.

\section{Hill function and its fluctuations-induced correction} \label{section4}
\begin{figure} [h!t]
  \centering
\includegraphics[width=.45\textwidth]{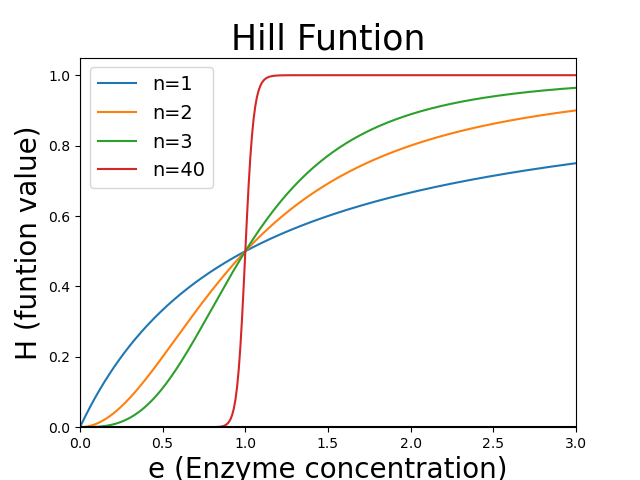}
  \caption{Deterministic Hill function, with $K=1$.}
  \label{fig.1}
\end{figure}

The Hill function is widely used in systems in which a ligand-receptor reaction occurs, or in our case, to capture the transcription rate of factors affecting mRNA synthesis. Generally, transcription factors act as activators or suppressors. In the case of an activating factor, the following Hill function is usually used,
\begin{align}
    H= \frac{\hat{e}^n}{K^n+ \hat{e}^n}. \label{3.9}
\end{align}
Here, $\hat{e}$ are the concentrations of the factors and $n$ is known as the Hill coefficient. Figure \ref{fig.1} shows how the Hill function behaves as an activator for various values of $n$. 
In the case of a repressor, the Hill function has the form $D=K^{n}/(K^{n}+\hat{e})$. 

Hill functions are derived by analyzing the deterministic dynamics of ligand-receptor reactions, and their use is widely extended, even in stochastic descriptions. However, its direct use as a transcription rate in the master equation of a stochastic system is an approximation \cite{Thomas,  Hole}. When studying stochastic models where Hill functions are used, one must be careful of the conditions under which they are employed \cite{Kim}. In this section, we derive the exact transcription rate owing to stationary ligand-receptor reactions by setting up the master equation of the entire reaction network of the system. This procedure allowed us to find corrections of the Hill function to consider the intrinsic fluctuations of ligand-receptor reactions. First, we analyze the Toggle switch as a particular case, and then derive a general expression for the corrections. 

\subsection{Toggle switch}
 We now analyze the Toggle switch genetic regulatory network to demonstrate a common approach used in the literature. We then pose the master equation of the complete reaction network corresponding to the system. A systematic treatment of the master equation will allow us to show that the commonly used Hill function is the first approximation and that corrections can be made.

The standard approach to describe the Toggle switch is to use the following chemical network 
{\small
\begin{align}
    \emptyset & \xrightarrow{\alpha D_d(R_2)} R_{1}, &
   \emptyset\xrightarrow{\alpha D_d(R_1) } R_{2}, \nonumber \\
    R_{1}& \xrightarrow{\beta} \emptyset, &
    R_{2}\xrightarrow{\beta} \emptyset,  
\end{align}}
where $D_{d}$ is the deterministic derived Hill function $D_d$, defined by
 \begin{align}
     D_d(x)=\frac{ (\frac{x}{K_{R}})^{n}}{1+(\frac{x}{K_{R}})^{n}}. \label{3.11}
 \end{align}
The corresponding master equation of the system is given by \cite{Scot, Sauro}

{\small
\begin{align}
  \frac{d}{dt} P(r_1,r_2,t)&= \alpha \Omega D_d\left(\frac{r_2}{\Omega} \right) P(r_1-1,r_2,t)- \beta {r_1} P(r_1,r_2,t)  \nonumber \\
  +& \beta ({r_1+1}) P(r_1+1,r_2,t) - \alpha \Omega D_d \left(\frac{r_2}{\Omega} \right) P(r_1,r_2,t) \nonumber \\
    &+ \alpha \Omega D_d \left(\frac{r_1}{\Omega} \right) P(r_1,r_2-1,t)- \beta {r_2} P(r_1,r_2,t) \nonumber \\
    +& \beta ({r_2+1}) P(r_1,r_2+1,t) - \alpha \Omega D_d \left(\frac{r_1}{\Omega} \right) P(r_1,r_2,t). \label{ClassicToggleSwitch}
\end{align}}
In the previous description of the Toggle switch, the Hill Function in the master equation (\ref{ClassicToggleSwitch}) was introduced under the assumption that the interaction of transcription factors with the promoter region, leading to the production of $R_{1}$ and $R_{2}$ implicitly involves the presence of transcription factors in the promoter region at a fixed concentration \cite{Thomas}. However, for a more precise representation, it is crucial to consider the inherent fluctuations in available active molecules, even at constant transcription factor concentrations. Therefore, we included the binding reactions between $R_{1}$ and $R_{2}$ to the corresponding transcription factors to accommodate these fluctuations in factor activity. Consequently, the reaction network of the Toggle Switch can be expressed as

{\small
\begin{align}
    P_{R_1} & +n R_{2}\stackbin[k_{-}]{k_{+}}{\rightleftarrows} P_{R_1}^{*}, &
    P_{R_2} +n R_{1}\stackbin[k_{-}]{k_{+}}{\rightleftarrows} P_{R_2}^{*}, \nonumber \\
    P_{R_{1}} & \xrightarrow{\alpha'} R_{1}, &
    P_{R_{2}}\xrightarrow{\alpha'} R_{2}, \nonumber \\
    R_{1}& \xrightarrow{\beta} \emptyset, &
    R_{2}\xrightarrow{\beta} \emptyset.  \label{4.35}
\end{align}}
We denote $r_1$ as the number of molecules of the chemical species $R_1$ and $p_1$ for active polymerase $P_{R_1}$ and $p_1^{*}$ for deactivated polymerase $P_{R_{1}}^{*}$. Similarly, $r_2$ represents the number of molecules of species $R_{2}$ and $p_2$ and $p_{2}^{*}$ are the corresponding active and inactivated polymerases $P_{R_2}$ and $P_{R_{2}}^{*}$. To clarify the first reactions, we must remember that in the Toggle Switch, $R_1$ inhibits $R_{2}$ and vice versa. Thus, if $n$ molecules of $R_{2}$ bind to the promoter region of $P_{R_1}$, they will deactivate it. The same process holds for bidding $R_{1}$ and $P_{R_{2}}$. Intermediate reactions can be used for $n > 1$ as in  \cite{Weiss}; however, in our study, we use the reactions in (\ref{4.35}).

If the reactions that occur in the promoter are fast (the first line of reactions in (\ref{4.35})), we consider that they have already reached an equilibrium. However, the rate of protein synthesis is slow. Therefore, we separated the master equation into a stationary part corresponding to fast reactions, and a dynamic part describing slow reactions \cite{Uri, Gout, Kim2}. 

{\tiny
\begin{align}
    \dot P(r_1,&r_2,p_1,p_2,p_1^{*},p_2^{*},t) = \alpha' ({p_2+1}) P(r_1-1,r_2,p_1,p_2,p_1^{*},p_2^{*},t)- \alpha' {p_1} P(r_1,r_2,p_1,p_2,p_1^{*},p_2^{*},t) \nonumber \\
    &+ \alpha' ({p_1+1}) P(r_1,r_2-1,p_1,p_2,p_1^{*},p_2^{*},t) - \alpha' {p_2} P(r_1,r_2,p_1,p_2,p_1^{*},p_2^{*},t) \nonumber \\
    &+ \beta ({r_1+1}) P(r_1+1,r_2,p_1,p_2,p_1^{*},p_2^{*},t)- \beta {r_1} P(r_1,r_2,p_1,p_2,p_1^{*},p_2^{*},t) \nonumber \\
    &+ \beta ({r_2+1}) P(r_1,r_2+1,p_1,p_2,p_1^{*},p_2^{*},t) - \beta {r_2} P(r_1,r_2,p_1,p_2,p_1^{*},p_2^{*},t), \nonumber \\
0=&   \Omega^n k_{-}(p_1^{*}+1) P(r_1,r_2-n,p_1-1,p_2,p_1^{*}+1,p_2^{*})- k_{+}p_1 \frac{r_2!}{(r_2-n)!} P(r_1,r_2,p_1,p_2,p_1^{*},p_2^{*})   \nonumber \\
    &+ k_{+}(p_1+1) \frac{(r_2+n)!}{r_2!} P(r_1,r_2+n,p_1+1,p_2,p_1^{*},p_2^{*}-1) - \Omega^n k_{-} p_1^{*} P(r_1,r_2,p_1,p_2,p_1^{*},p_2^{*})  \nonumber \\
    &+ \Omega^n k_{-}(p_2^{*}+1) P(r_1-n,r_2,p_1,p_2-1,p_1^{*},p_2^{*}+1) - k_{+}p_2 \frac{r_1!}{(r_1-n)!} P(r_1,r_2,p_1,p_2,p_1^{*},p_2^{*})   \nonumber \\
    &+ k_{+}(p_2+1) \frac{(r_1+n)!}{r_1!} P(r_1+n,r_2,p_1,p_2+1,p_1^{*},p_2^{*}-1) - \Omega^n k_{-} p_2^{*} P(r_1,r_2,p_1,p_2,p_1^{*},p_2^{*}), \label{4.38}
\end{align}}  
In these equations, we denote $r_1$ as the number of elements of the chemical species $R_1$ and $p_1$ for the active polymerase, $p_1^{*}$ for the deactivated polymerase, and we denote the remaining variables similarly.

We now take the average over the stationary variables $p_1$ and $p_2$ on the equation (\ref{4.38}) obtaining and effective master equation for only the variables $r_1$ and $r_2$,

{\small
\begin{align}
    \dot P(r_1,r_2,t)& = \alpha' \braket{p_2}_{r_{2}} P(r_1-1,r_2,t)- \alpha' \braket{p_1}_{r_{1}} P(r_1,r_2,t) \nonumber \\
    &+ \alpha' \braket{p_1}_{r_{1}} P(r_1,r_2-1,t) - \alpha' \braket{p_2}_{r_{2}} P(r_1,r_2,t) \nonumber \\
    &+ \beta ({r_1+1}) P(r_1+1,r_2,t)- \beta {r_1} P(r_1,r_2,t) \nonumber \\
    &+ \beta ({r_2+1}) P(r_1,r_2+1,t) - \beta {r_2} P(r_1,r_2,t).
\end{align}}
On the other hand, the solution of the stationary distribution (\ref{4.38}) can be explicitly obtained (see appendix \ref{A0}), and thus we can calculate 

\begin{equation}
    \braket{p_2}_{r_{2}} = A \left(  \frac{K^n}{K^n + \frac{\braket{\frac{r_{1}!}{(r_{2}-n)!}}_{r_2}}{\Omega^n }} \right) \approx A \left(\frac{K^n}{K^n + \frac{\braket{r_{2}}^n +  \frac{n}{2}(n-1) \sigma^2(r_{2},r_{2}) \braket{r_{2}}^{n-2} }{\Omega^n }} \right),
\end{equation}
where $K\equiv\frac{k_{+}}{k_{-}}$, $A$ is a constant and we have used the fact that
\begin{align}
    \braket{f(\mathbf{X})} \approx &   f(\braket{\mathbf{X}})  +  \sum_i \sum_j \frac{\sigma^2(X_i,X_j)}{2} \left[ \frac{\partial^2 f(\mathbf{X})}{\partial X_i \partial X_j}\right]_{\mathbf{X}=\braket{\mathbf{X}}}. \label{4.42} 
\end{align}
We also find a similar expression for $ \braket{p_1}_{r_{1}}$ and 
thus, if we define 

\begin{align}
    D_1(\braket{x}, \sigma^2(x,x),\Omega) \equiv \frac{K^{n}}{K^{n} + \frac{\braket{x}^n +  \frac{n}{2}(n-1) \sigma^2(x,x) \braket{x}^{n-2} }{\Omega^n }}. \label{4.43}
\end{align} 

We end up with a closed master equation describing $r_{1}$ and $r_{2}$

{\footnotesize
\begin{align}
    \dot P(r_1,r_2,t)&= \alpha \Omega D_1(\braket{r_1}, \sigma^2(r_1,r_1),{\Omega})P(r_1,r_2-1,t)- \alpha \Omega  D_1(\braket{r_2}, \sigma^2(r_2,r_2),{\Omega}) P(r_1,r_2,t) \nonumber \\
    &+ \alpha \Omega D_1(\braket{r_2}, \sigma^2(r_2,r_2),{\Omega}) P(r_1-1,r_2,t) - \alpha \Omega D_1(\braket{r_1}, \sigma^2(r_1,r_1),{\Omega}) P(r_1,r_2,t) \nonumber \\
    &+ \beta (r_1+1) P(r_1+1,r_2,t)- \beta {r_1} P(r_1,r_2,t) \nonumber \\
    &+ \beta (r_2+1) P(r_1,r_2+1,t) - \beta {r_2} P(r_1,r_2,t). \label{3.15}
\end{align}}

(We have substituted $\alpha'= \frac{\alpha \Omega}{A}$).
Function $D_{1}$ plays the role of the Hill function; thus, it can be seen as the first correction of the Hill function due to fluctuations in species concentrations. 

In the following, we refer to $D_{1}$ defined in (\ref{4.43})  as the Hill function with stochastic corrections.

We then use the master equation (\ref{3.15}) to obtain the MFK approximation (see section 2). Thus the set of differential equations describing the evolution of concentrations $\hat{r}_1\equiv \frac{\braket{r_1}}{\Omega}$ in the Toggle switch are
\begin{align}
    \frac{\partial \hat{r}_1}{\partial t} =& \alpha D_1(\hat{r}_2, \sigma^2(\hat{r}_2,\hat{r}_2))- \beta \hat{r}_1, \nonumber \\
    \frac{\partial \sigma^2(\hat{r}_1, \hat{r}_1)}{\partial t}=& \frac{\alpha D_1(\hat{r}_2, \sigma^2(\hat{r}_2,\hat{r}_2))+ \beta \hat{r}_1}{\Omega} - 2 \beta \sigma^2(\hat{r}_1, \hat{r}_1)
\end{align}
 A similar result was obtained for the dynamics of $\hat{r}_{2}$.

From this example, it is clear that directly introducing the Hill function into the master equation does not consider fluctuations ($\sigma^{2}(x,x)=0$). Furthermore, we found that making the following transformation
\begin{align}
    \hat{x}^n  \rightarrow \hat{x}^n +  \frac{n}{2}(n-1) \sigma^2(\hat{x},\hat{x}) \hat{x}^{n-2}
\end{align}
in the evaluation of a Hill function is sufficient either in the case of activators (\ref{3.9}) or in repressors (\ref{3.11}). 
 It is also possible to make this substitution even for other types of Hill functions, such as those shown in Appendix \ref{C1}.

In the following section, we employ the approximation presented in Section \ref{section2}, as well as the Hill function with stochastic corrections, to analyze some particular cases.

\section{Stochastic genetic regulation networks} \label{section5}
Because we are interested in describing the stochastic version of a gene regulatory network, we briefly explain how a transcription/translation module (TTM) works. When several of these modules are connected, they form a gene regulatory network. Strictly speaking, one of the proteins produced by one of these modules acts as a transcription factor for other TTMs. 

\begin{figure} [h!t]
  \centering
\includegraphics[width=0.45\textwidth]{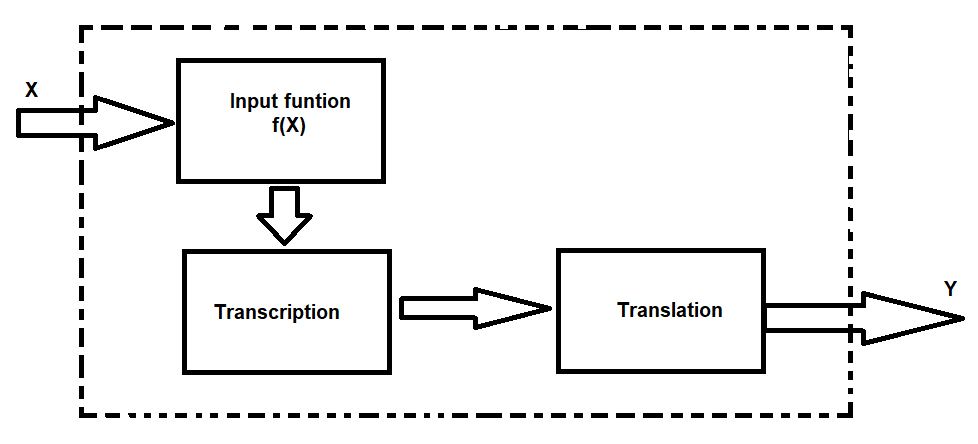}
  \caption{Transcription/translation module.}
  \label{fig.2}
\end{figure}

Figure \ref{fig.2} shows a simplified schematic view of a TTM, where a $X$ protein is the input of the module acting as a transcription factor, the transcription process occurs when the mRNA is synthesized, and translation occurs when some new $Y$ proteins are synthesized from the mRNA.

Hill function is typically used to express the production rate of a protein in terms of its activator (repressor). More specifically, it helps describe the number of active promoters as a function of transcription factor concentration. 

\subsection{Stochastic genetic regulatory network }

The most simple description of a MTT can be described with the following reactions
\begin{align}
        x & \stackbin{H}{\rightarrow} \text{mRNA} \stackbin{k_2}{\rightarrow} y & \nonumber \\
        \text{mRNA} &\stackbin{\gamma_m}{\rightarrow} 0, \nonumber \\
        \text{y} &  \stackbin{\gamma_p}{\rightarrow} 0. \label{5.33}
\end{align}
The first reaction indicates that mRNA is synthesized by transcription factor $x$ with a reaction rate proportional to the Hill function. Subsequently, it is synthesized into a protein $y$, and the last two reactions indicate that the mRNA and protein are degraded and diluted. In the reaction rates, $H$ is a Hill function, which can be an activator or a suppressor depending on the system. Hill function would be a repressor if the protein suppressed gene activation; otherwise, it would be an activator. 

In a more general sense, many transcription factors can act on the same gene giving effectively a more complex Hill function of the form
\begin{align}
    H_{i}(\mathbf{p})= \frac{ \sum_j \lvert A_{ij}\lvert \left( \frac{p_j}{\Omega K_j}\right)^{n_{ij} * A_{ij}}}{1 + \sum_j \lvert A_{ij} \lvert \left( \frac{p_j}{\Omega K_j}\right)^{n_{ij} * A_{ij}}},
\end{align}
(For more details on the derivation of this equation, please refer to Appendix \ref{B1}), where $A_{ij}$ is the connection matrix with elements $A_{il}=1$ if $l$ activates $i$ and $A_{ij}=-1$ if $j$ represents $i$ otherwise, $A_{ij}=0$. The numbers $n_{ij}$ are called the Hill coefficients and $K_{j}$ are the dissociation constants.

The general expression of the master equation describing a gene regulatory network is given by (\ref{5.33}), 
{\small
\begin{align}
    \frac{\partial P( \mathbf{m}, \mathbf{p},t)}{\partial t}=&  \sum_i \bigg(   { \Omega k_{1i}} \left(   H_{i}(\mathbf{p},{\Omega})P(\mathbf{m},\mathbf{p},m_i-1,t)-  H_{i}(\mathbf{p},{\Omega})P(\mathbf{m}, \mathbf{p},t)\right)  \bigg. \nonumber \\
    &+   \frac{m_i+1}{ \tau_{1i}}P(\mathbf{m},\mathbf{p},m_i+1,t) -\frac{m_i}{ \tau_{1i}}P(\mathbf{m}, \mathbf{p},t) \nonumber \\
    &+  k_{2i} {m_i} P(\mathbf{m},\mathbf{p},p_i-1,t)- k_{2i} {m_i} P(\mathbf{m}, \mathbf{p},t) \nonumber \\
    & \left.  + \frac{p_i+1}{\tau_{2i}}P(\mathbf{m},\mathbf{p},p_i+1,t) -\frac{p_i}{ \tau_{2i}}P(\mathbf{m}, \mathbf{p},t)\right) , \label{5.52}
\end{align}}
where $\tau_{1i}= \frac{1}{\gamma_{m_i}}$ and $\tau_{2i}= \frac{1}{\gamma_{p_i}}$ are mRNA and protein degradation times, respectively.   Index $i$ labels each MTT, and there is a sum over all the network modules. In this equation, an extra argument is added (e.g., $m_i-1$) to emphasize which variable changes depending on $i$.

In expression (\ref{5.52}), it is customary to use operators to supply the Hill function \cite{RS, TSS} or directly use the deterministic Hill function \cite{Scot}; however, we have the option of using a generalized Hill function with stochastic corrections.

From equation (\ref{5.52}), we calculate the steady-state distribution of the system, from which we obtain
{\footnotesize
\begin{align}
        P_{ss}(\mathbf{m}, \mathbf{p})=&   \prod_i \frac{(k_{1i} \tau_{1i} \Omega \braket{H_i(p_l, \Omega)})^{m_i} e ^{-k_{1i} \tau_{1i}  \Omega \braket{H_i(p_l, \Omega)} }}{m_i!} \frac{(k_{2i} \tau_{2i} \braket{m_i})^{p_i} e ^{-k_{2i} \tau_{2i} \braket{m_i}}}{p_i!}. \label{5.26}
\end{align}}
If we use the Hill function with stochastic corrections and take limits, as in  \cite{Hole}, it is possible to reproduce the steady-state distributions calculated for nonheuristic stochastic models.

From eq. (\ref{5.26}), the fluctuations in the steady state can been exactly calculated to be
\begin{align}
       \eta_{m_i}^2&= \frac{\braket{m_i}}{\Omega^2}, &
        \eta_{p_i}^2&= \frac{\braket{p_i}}{\Omega^2}.  \label{fluctuations}
\end{align}
This particular expression, which is proportional to the first moment of the variables, originates from the fact that a Poisson distribution purely describes the stationary distribution.
In this case, $\braket{m_i}$ and $\braket{p_i}$ are the average number of molecules, but in terms of concentration, the fluctuations are proportional to the inverse of the system size in contrast to its square, as in (\ref{fluctuations}).

Two examples are presented to illustrate the proposed analysis and highlight its benefits.

 \subsection{Repressilator}
 
  \begin{figure} [h!t]
  \centering
\includegraphics[width=0.25\textwidth]{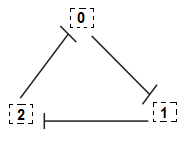}
  \caption{Represilator. Each node represents a TTM.}
  \label{fig.4}
\end{figure}

Figure \ref{fig.4} shows a graphical representation of the repressilator. This simple but important system was developed experimentally \cite{Elo}. With the help of figure \ref{fig.4} we build its connection matrix $A$
  \begin{align}
     A_{ij}= \begin{pmatrix}
		0&0 & -1 \\
		-1& 0 & 0 \\
		0 & -1 & 0
	\end{pmatrix},
 \end{align}
 where the first row of matrix $A$ indicates that a protein synthesized in module 3 acts as a suppressor in module 1 and similarly in the other modules. The corresponding master equation is 
 {\footnotesize
\begin{align}
    \frac{\partial P(\mathbf{m},\mathbf{p},t)}{\partial t}=& \sum_{i=1}^{3} \bigg(  \left(   k_{1}^{-} \Omega \left( H_i(\mathbf{p},{\Omega})P(\mathbf{m},\mathbf{p},m_i-1,t)-H_i(\mathbf{p},{\Omega}) P(\mathbf{m},\mathbf{p},t) \right)  \right) \bigg.  \nonumber \\
    &+   \frac{m_i+1}{ \tau_{1i}}P(\mathbf{m},\mathbf{p},m_i+1,t) -\frac{m_i}{ \tau_{1i}}P(\mathbf{m},\mathbf{p},t) \nonumber \\
    &+  k_{2i} {m_i} P(\mathbf{m},\mathbf{p},p_i-1,t)- k_{2i} {m_i} P(\mathbf{m},\mathbf{p},t) \nonumber \\
    & \left.  + \frac{p_i+1}{\tau_{2i}}P(\mathbf{m},\mathbf{p},p_i+1,t) -\frac{p_i}{ \tau_{2i}}P(\mathbf{m},\mathbf{p},t)\right) . 
\end{align}}
The Hill functions $H_i(\mathbf{p},{\Omega})$ are repressors. 

\begin{figure} [h!t]
\centering
\includegraphics[width=0.45\textwidth]{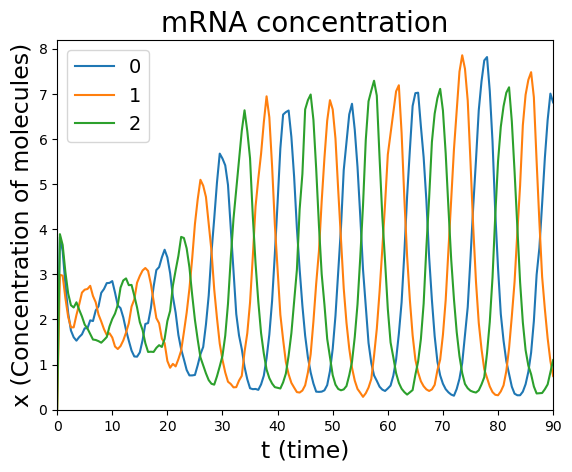}
\includegraphics[width=0.45\textwidth]{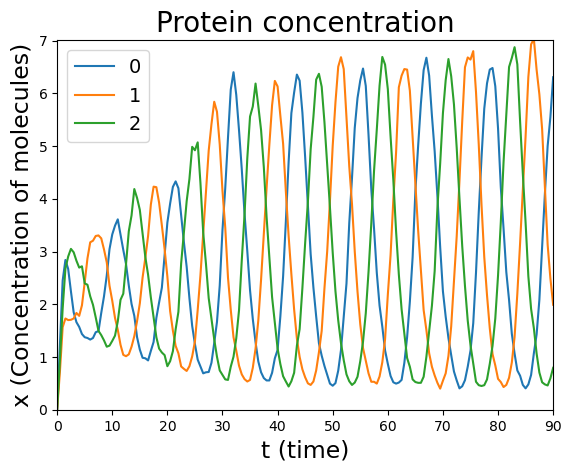}
  \caption{Repressilator. Stochastic simulation using the Gillespie algorithm and deterministic Hill function. The first figure shows the concentration of mRNA and the second figure shows the concentration of proteins. The following parameters are used:  $n=2$, $\Omega=200$, $\tau_{1i}=\tau_{2i}=1$, $k_{1i}=k_{2i}= 10$.}
  \label{fig.5}
\end{figure}

We can now analyze the system using three different approaches. First, we simulated the system dynamics using the Gillespie algorithm, assuming deterministic Hill functions for reaction rates. Second, the description can be performed by Mass Fluctuation Kinetics (MFK) using the deterministically derived Hill function. Finally, we can describe the system dynamics with MFK again, but using the Hill function with stochastic correction, an expression similar to (\ref{4.43}).

\begin{figure} [h!t]
  \begin{subfigure}{\linewidth}
\includegraphics[width=.3\textwidth]{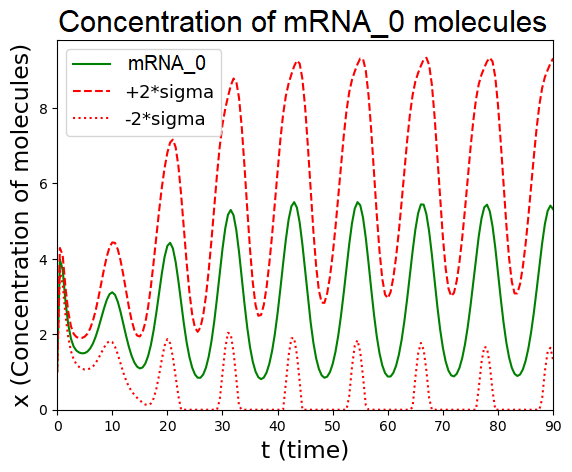} \hfill
\includegraphics[width=.3\textwidth]{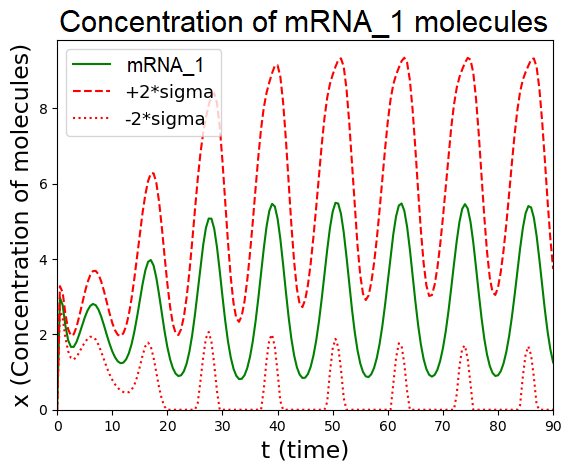} \hfill
\includegraphics[width=.3\textwidth]{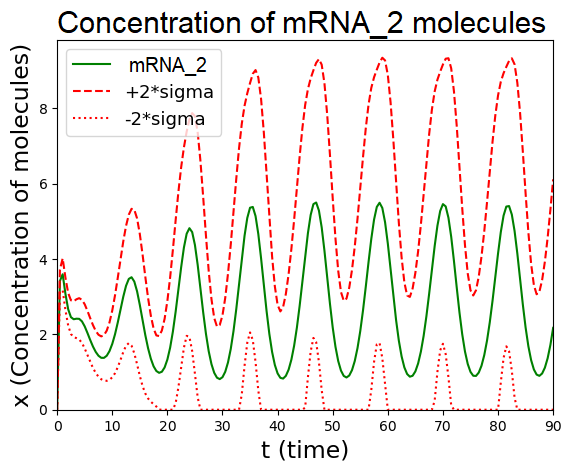}
  \end{subfigure}\par\medskip
 \begin{subfigure}{\linewidth}
 \centering
\includegraphics[width=.3\textwidth]{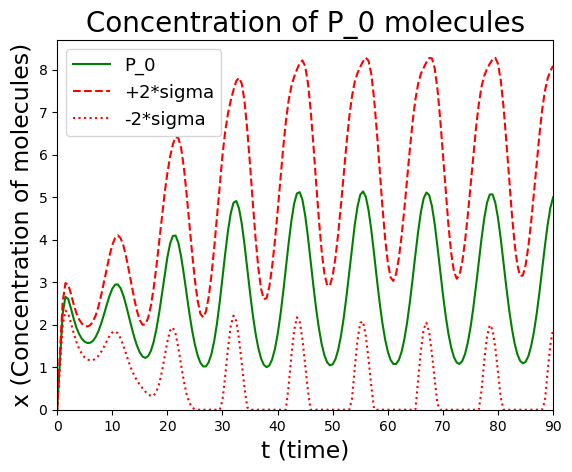} \hfill
\includegraphics[width=.3\textwidth]{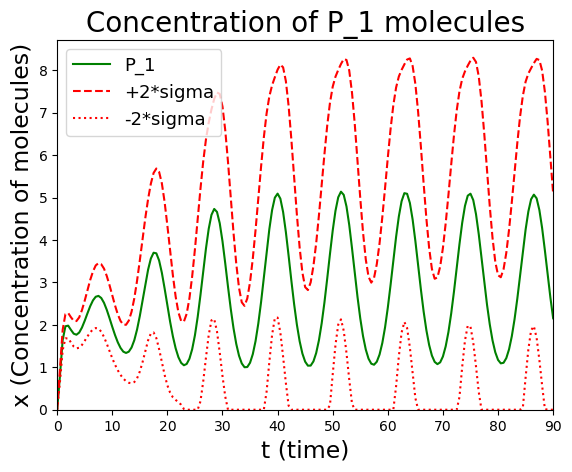}\hfill
\includegraphics[width=.3\textwidth]{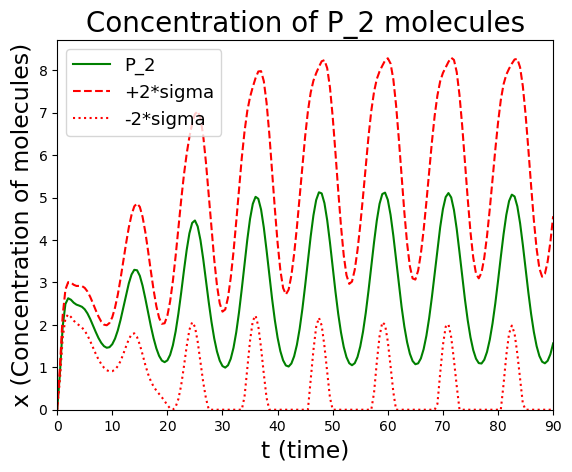}
  \end{subfigure}
\caption{Repressilator. Dynamics of deterministic concentrations and fluctuations from stochastic simulations with an assembly of $10000$ simulations. In the upper figures, the dynamics of mRNAs are observed in the lower ones of the proteins. The green lines represent the deterministic concentrations and the red dotted lines represent the bands where fluctuations were observed. The following parameters were used: $Arm_1(0)=Arm_2(0)=r_0(0)=r_1(0)=r_2(0)=0$, $Arm_1(0)= 1$, n=2, $\Omega=200$, $\tau_{1i}=\tau_{2i}=1$, $k_{1i}=k_{2i}= 10$.}
  \label{fig.6}
\end{figure}

The Gillespie simulation of the system is shown in figures \ref{fig.5}. In this case, we used the deterministic Hill function, and the size of the system was 200 (as in the Elowitz experiment \cite{Elo}). Protein and mRNA concentrations are plotted. This is simply a realization of a stochastic system; thus, the oscillation amplitudes are not uniform. 

By creating an ensemble of simulations, the average or deterministic dynamics and fluctuations were obtained (see Fig. \ref{fig.6}).
In Fig. \ref{fig.6} it is plotted the concentration of the mRNA and the region of the fluctuations. 

\begin{figure} [h!t]
  \begin{subfigure}{\linewidth}
\includegraphics[width=.3\textwidth]{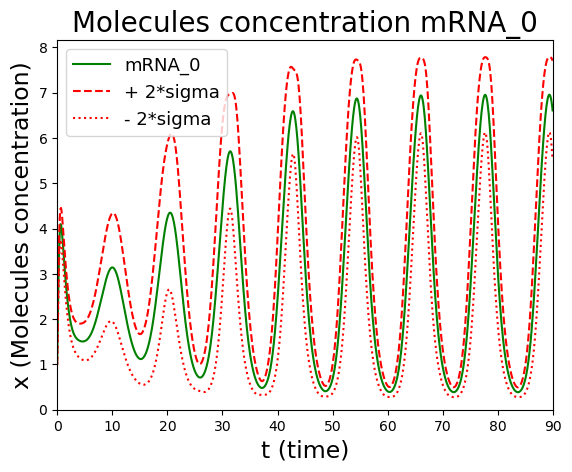}\hfill
\includegraphics[width=.3\textwidth]{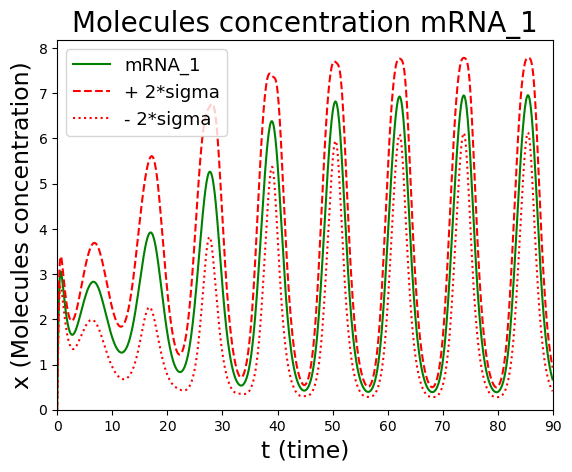}\hfill
\includegraphics[width=.3\textwidth]{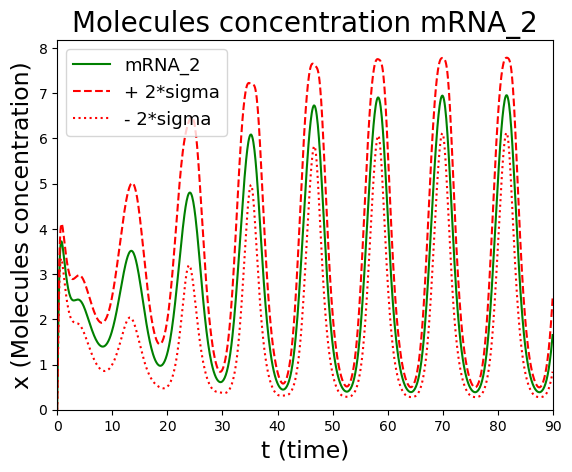}
  \end{subfigure}\par\medskip
 \begin{subfigure}{\linewidth}
 \centering
\includegraphics[width=.3\textwidth]{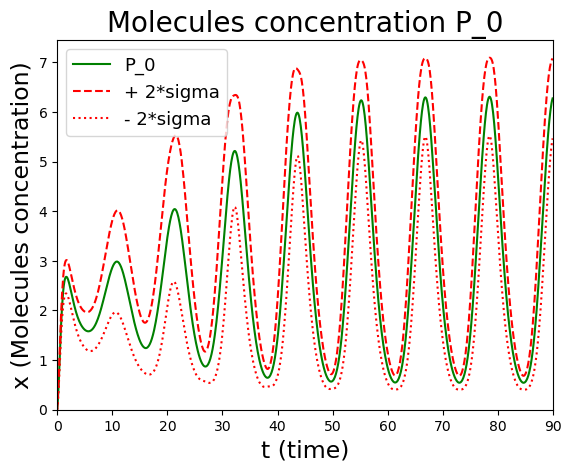}\hfill
\includegraphics[width=.3\textwidth]{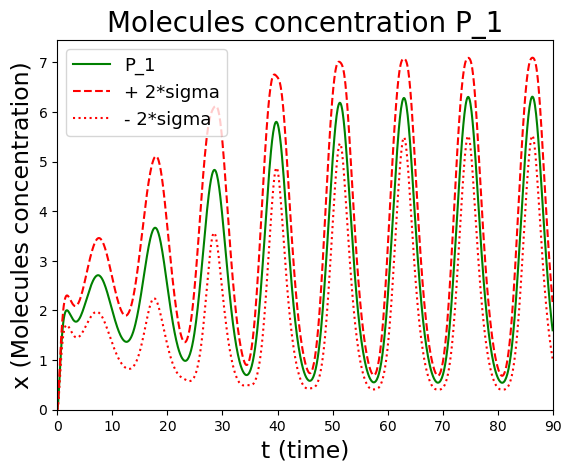}\hfill
\includegraphics[width=.3\textwidth]{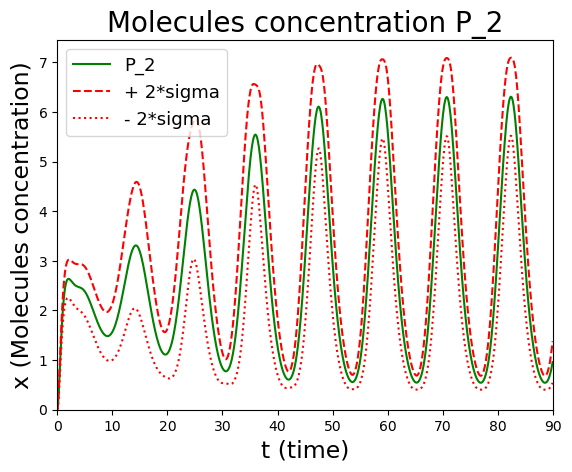}
\caption{Deterministic Hill function.}
\end{subfigure} \par\medskip
  \begin{subfigure}{\linewidth}
\includegraphics[width=.3\textwidth]{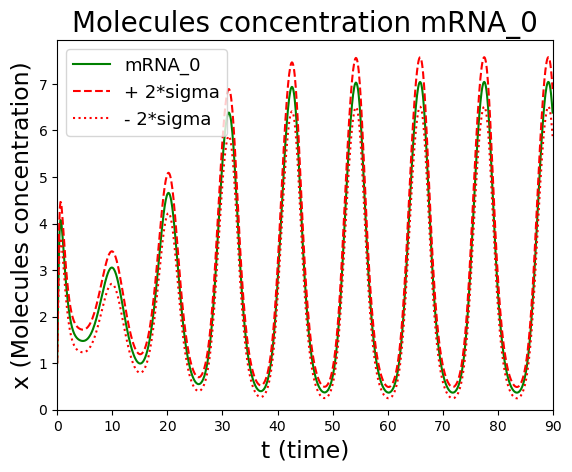}\hfill
\includegraphics[width=.3\textwidth]{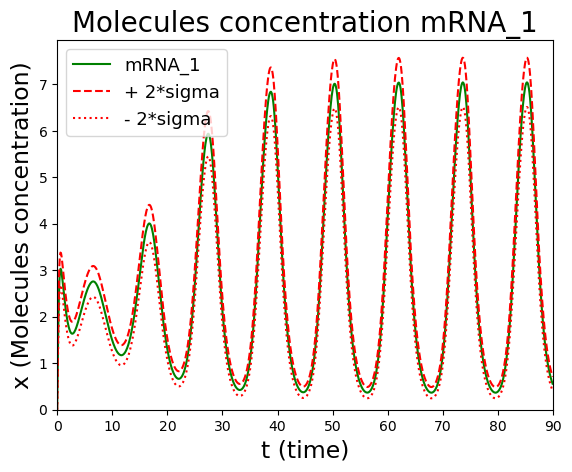}\hfill
\includegraphics[width=.3\textwidth]{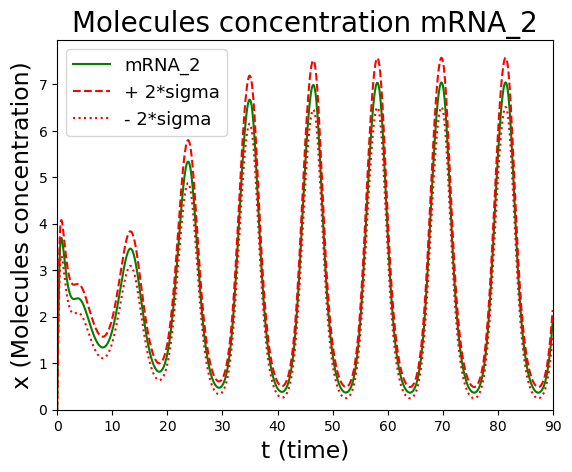}
  \end{subfigure}\par\medskip
 \begin{subfigure}{\linewidth}
 \centering
\includegraphics[width=.3\textwidth]{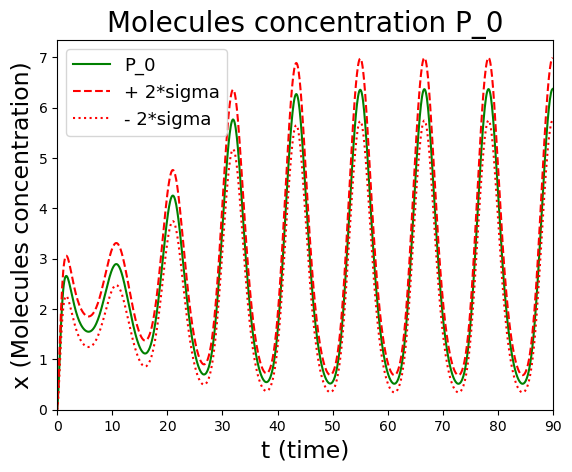}\hfill
\includegraphics[width=.3\textwidth]{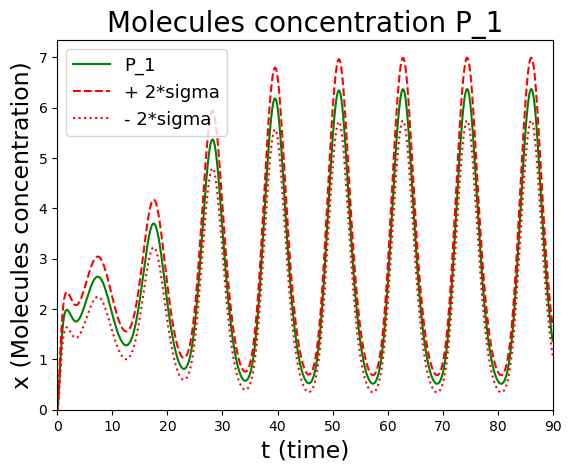}\hfill
\includegraphics[width=.3\textwidth]{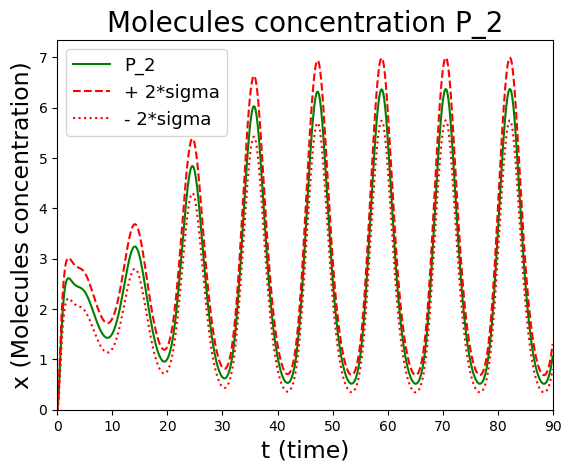}
\caption{ Hill function with stochastic corrections.}
\end{subfigure}
\caption{Represilador. The dynamics and fluctuations of the system were modeled using the MFK approach, as shown in Figure (a). A deterministic Hill function was used. In (b), the Hill function with stochastic corrections was used. For both (a) and (b), the figures at the top show the dynamics of mRNAs, whereas those at the bottom show the dynamics of the proteins. The green lines represent the deterministic concentrations, and the red dotted lines represent the bands where fluctuations were observed. The following parameters have been used: $Arm_2(0)=r_1(0)=r_2(0)=0$, $Arm_1(0)= 1$, n=2, $\Omega=200$, $\tau_{2i}=\tau_{3i}=1$, $k_{2i}=k_{3i}= 10$.}
  \label{fig.7}
\end{figure}

In figure \ref{fig.7}(a), we plot the analytically obtained deterministic dynamics and deduce the region of fluctuation using  MFK.

 Finally, using the Hill function with stochastic correction, we obtained Figure \ref{fig.7}(b). The most notable result is that the size of the fluctuations is considerably reduced compared with Figures \ref{fig.6} and \ref{fig.7}(a). At first sight, this might be counterintuitive, but it shows that the stochastic effect on the reaction rates might have no linear effects on gene regulatory networks that make them robust with respect to intrinsic fluctuation.

\begin{figure} [h!t]
  \centering
\includegraphics[width=0.45\textwidth]{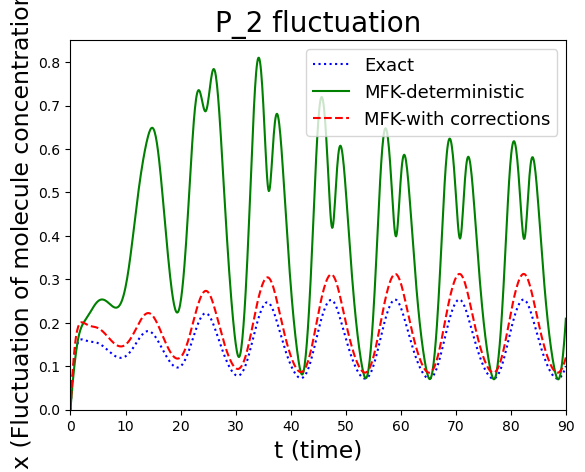}
  \caption{Fluctuations. We plotted the fluctuations obtained using the MFK approach first with the deterministic Hill function (green line) and then with the Hill function with stochastic corrections (red line). Finally, we compared these results with those obtained analytically in steady state (blue line).} \label{fig.8}
\end{figure}

We can compare the predicted fluctuations in the steady state of the approximations with the exact result. In fig. \ref{fig.8}, we plotted the characteristic size of the fluctuations using the deterministic Hill function and the Hill function with stochastic corrections, and it is compared with the exact calculation. It is observed that the corrected Hill function provides a much better description of the stochastic system.

\subsection{Activator-repressor clock}

  \begin{figure} [!ht]
  \centering
\includegraphics[width=0.25\textwidth]{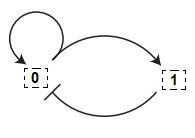}
  \caption{Activador-represor clock. Each node represents a TTM.}
  \label{fig.9}
\end{figure}

A genetic network in which several transcription factors participate is the activator-repressor clock. First, we analyze the system with the Gillespie algorithm using the deterministic Hill function. Then, we use the MFK with the deterministic Hill function, and finally, we analyze it using stochastic corrections to the Hill function. Figure \ref{fig.9} presents a graphical representation of the model.  

In module 1, the protein synthesized by module 2 acts as a suppressor, whereas the protein synthesized by module 1 acts as an activator on itself. The protein synthesized by module 1 acts as an activator of module 2. We assume that $k_{1i}=k_{1}$, also $n_{11}=n_{12}=n_{21}=2$ and $n_{22}=0$, $\beta_{11}= \beta{21}=1$, $\beta_{11}= 0.0004$ so the master equation is

 {\footnotesize
\begin{align}
    \frac{\partial P(\mathbf{m},\mathbf{p},t)}{\partial t}&= \sum_{i=1}^{2} \left(  k_{1i} \Omega \left( H_g^{i}(\mathbf{p},{\Omega})P(\mathbf{m},\mathbf{p},m_i-1,t)-H_g^{i}(\mathbf{p},{\Omega}) P(\mathbf{m},\mathbf{p},t)  \right)\right.  \nonumber \\
    &+ \frac{m_i+1}{\tau_{2i}}P(\mathbf{m},\mathbf{p},m_i+1,t) -\frac{m_i}{\tau_{2i}}P(\mathbf{m},\mathbf{p},t) \nonumber \\
    &+   k_{3i} {m_i} P(\mathbf{m},\mathbf{p},p_i-1,t)- k_{3i} {m_i} P(\mathbf{m},\mathbf{p},t) \nonumber \\
    &+ \left. \frac{p_i+1}{ \tau_{3i}}P(\mathbf{m},\mathbf{p},p_i+1,t) -\frac{p_i}{ \tau_{3i}}P(\mathbf{m},\mathbf{p},t)\right) . 
\end{align}}
Where the Hill functions take the form
 {\small
 \begin{align}
     H_{g}^{1}({p_1},p_2,{\Omega})=&  \frac{ \beta_1 \left(\frac{p_1}{K_1' \Omega}\right)^{{n_{11}}} + \alpha_0 }{1 + \left(\frac{p_1}{K_1' \Omega}\right)^{{n_{11}}} + \left(\frac{p_2}{K_2 \Omega}\right)^{n_{12}} }, &
     H_{g}^{2}({p_1},{\Omega})=& \frac{ \beta_2 \left(\frac{p_1}{K_1 \Omega}\right)^{{n_{21}}}  }{1 + \left(\frac{p_1}{K_1 \Omega}\right)^{{n_{21}}} }.
 \end{align}}

\begin{figure} [h!t]
  \centering
\includegraphics[width=0.45\textwidth]{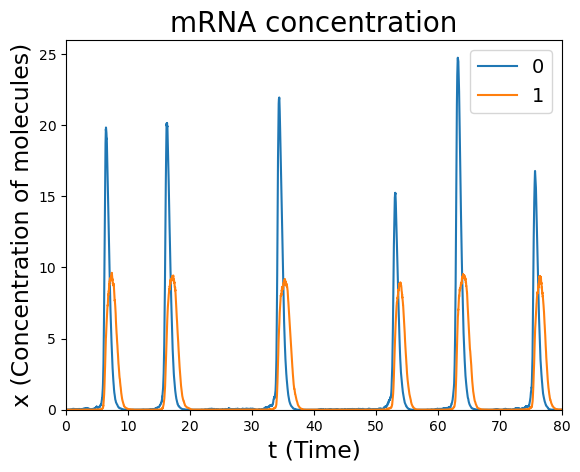}
\includegraphics[width=0.45\textwidth]{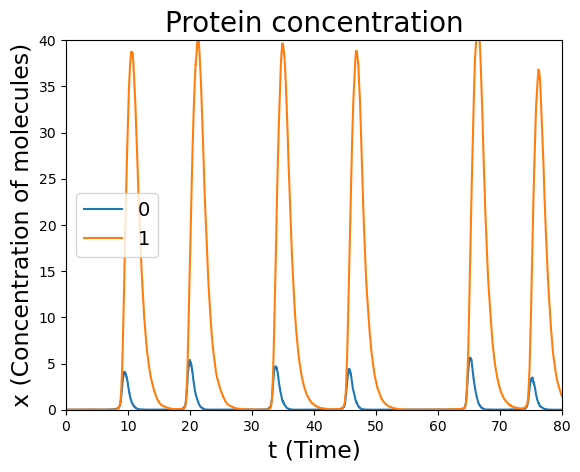}
  \caption{Activator-repressor clock. Stochastic simulation using the Gillespie algorithm and the deterministic Hill function. In the first figure, the concentration of mRNA is shown, and in the second figure, the concentration of proteins is shown. The following parameters have been used: $n_{11}=n_{12}=n_{21}=$2, $\Omega=180$, $k_{11}=250$, $k_{12}=30$, $\alpha_0= 0.0004$  $\tau_{2i}=\frac{1}{3}$, $\tau_{31}=\frac{1}{3}$, $\tau_{32}=1$, $k_{21}=1$, $k_{21}=3$.} \label{fig.10}
\end{figure}

By employing the Gillespie algorithm to simulate the system and using the deterministic Hill function, we obtain Figure \ref{fig.10}. 

\begin{figure} [h!t]
  \centering
\includegraphics[width=0.45\textwidth]{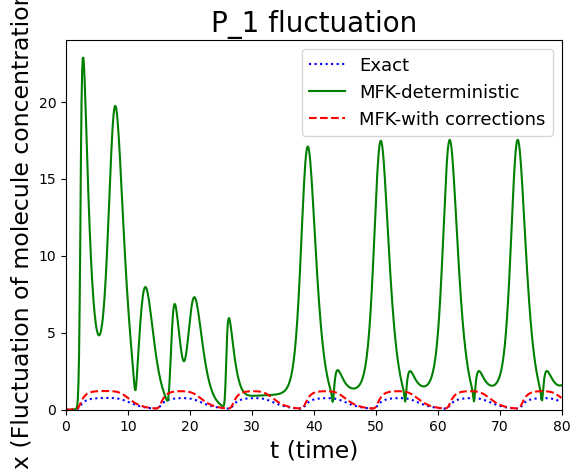}
\includegraphics[width=0.45\textwidth]{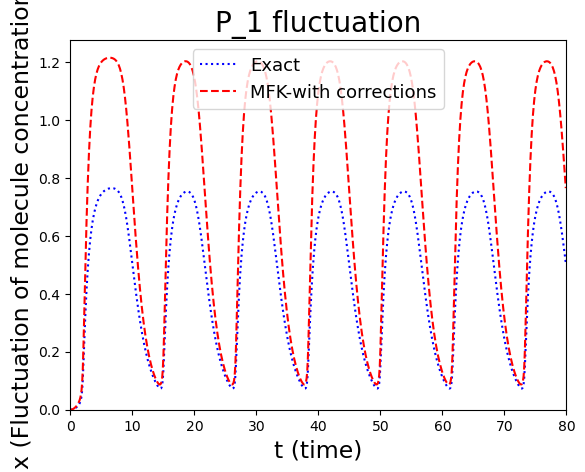}
  \caption{Fluctuations. We plotted the fluctuations obtained using the MFK approach, first with the deterministic Hill function (green line) and then with the Hill function with stochastic corrections (red line). Finally, we compared these results with those obtained analytically in the steady state (blue line).} \label{fig.13}
\end{figure}
  
\begin{figure} [h!t]
  \begin{subfigure}{\linewidth}
  \includegraphics[width=.24\textwidth]{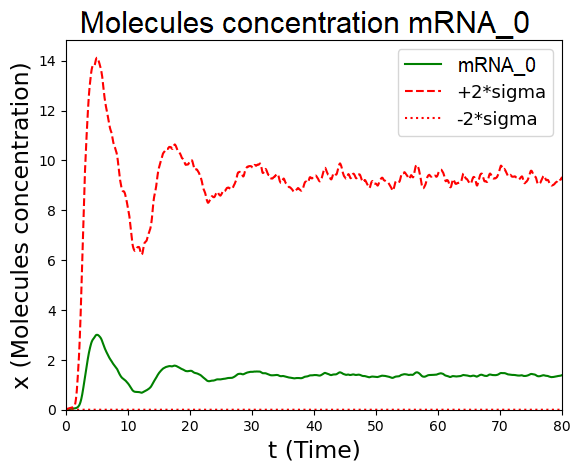}\hfill
  \includegraphics[width=.24\textwidth]{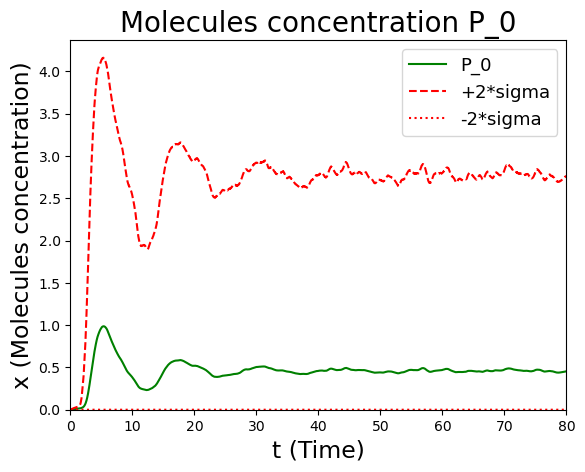} \hfill
\includegraphics[width=.24\textwidth]{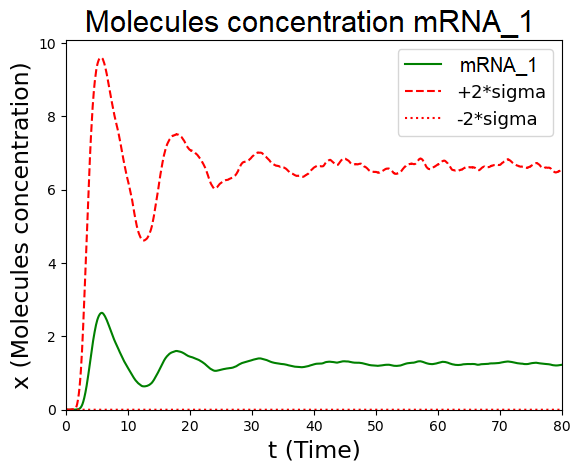}\hfill
\includegraphics[width=.24\textwidth]{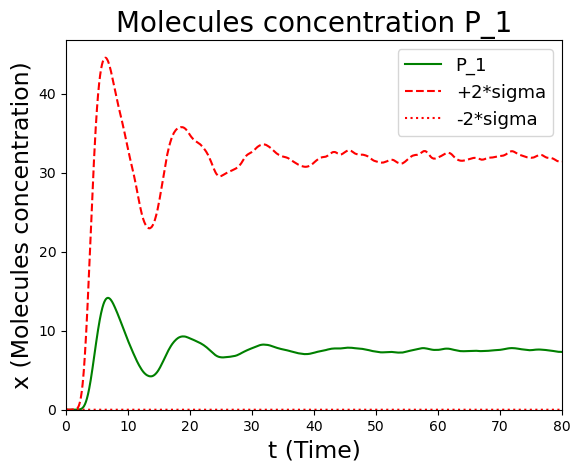}
\caption{Stochastic simulations.}
\end{subfigure}
  \begin{subfigure}{\linewidth}
\includegraphics[width=.24\textwidth]{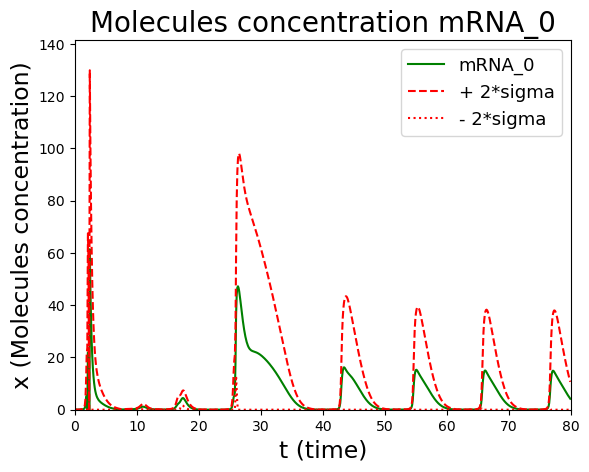}\hfill
\includegraphics[width=.24\textwidth]{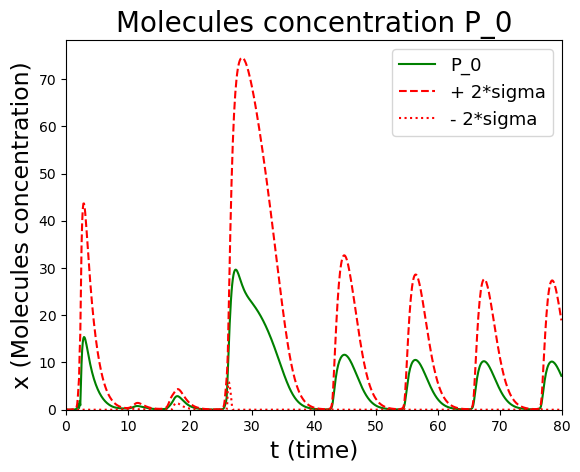}\hfill
\includegraphics[width=.24\textwidth]{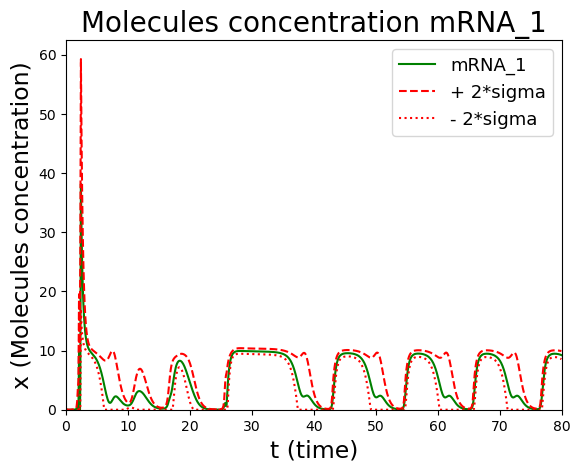}\hfill
\includegraphics[width=.24\textwidth]{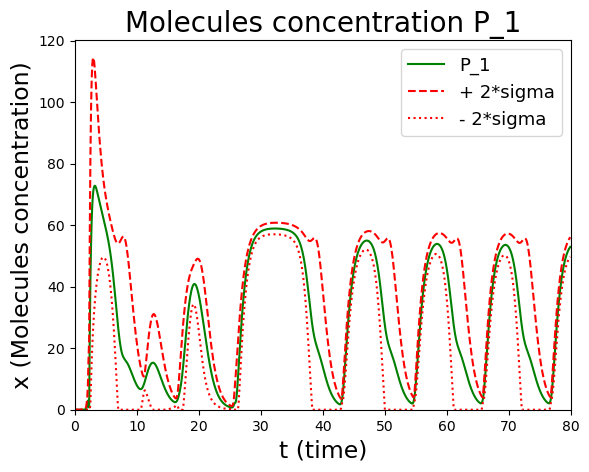}
\caption{Deterministic Hill function.}
\end{subfigure} \par\medskip
  \begin{subfigure}{\linewidth}
\includegraphics[width=.24\textwidth]{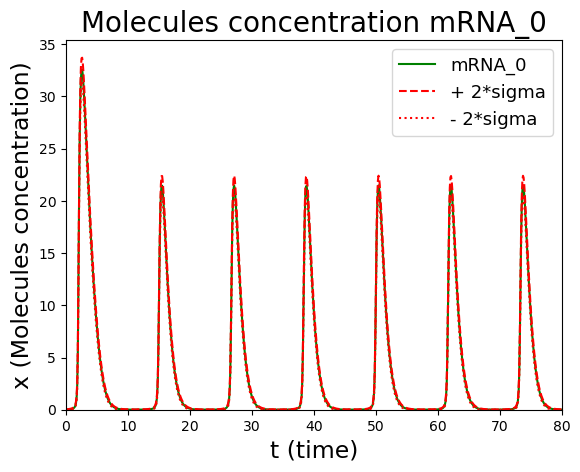}\hfill
\includegraphics[width=.24\textwidth]{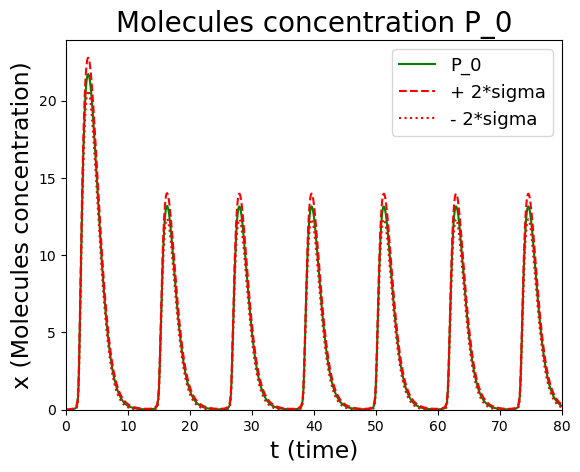}\hfill
\includegraphics[width=.24\textwidth]{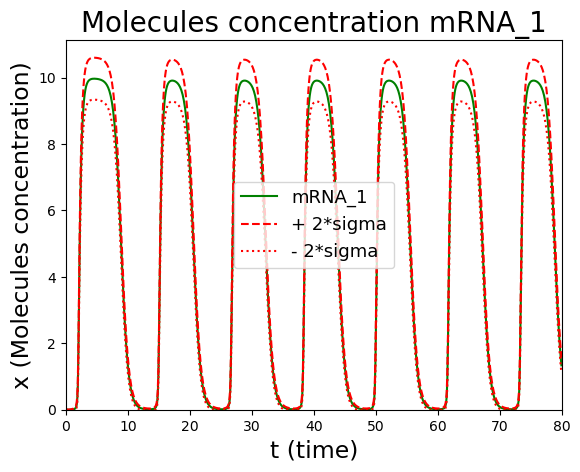}\hfill
\includegraphics[width=.24\textwidth]{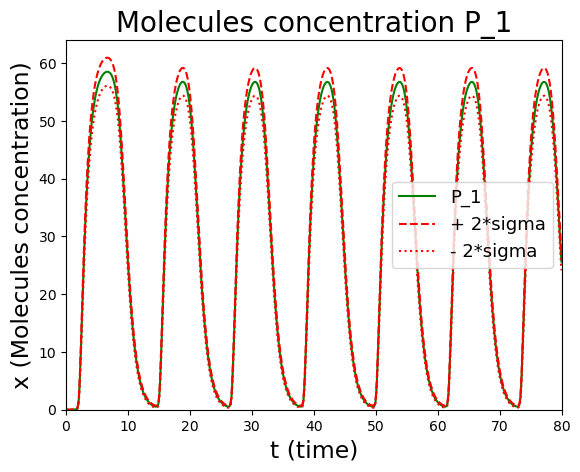}
\caption{Hill function with stochastic corrections.}
\end{subfigure}
  \caption{Activator-repressor clock. The dynamics and fluctuations of the system were modeled using stochastic simulations with an assembly size$=10000$, as shown in Figure (a),  the MFK approach (b), and the deterministic Hill function was used. In (c), the MFK approach and the Hill function with stochastic corrections are used.  The green lines represent the deterministic concentrations, and the red dotted lines represent the bands where fluctuations are found. The following parameters have been used:   $n_{11}=n_{12}=n_{21}=$2, $\Omega=180$, $k_{11}=250$, $k_{12}=30$, $\alpha_0= 0.0004$  $\tau_{2i}=\frac{1}{3}$, $\tau_{31}=\frac{1}{3}$, $\tau_{32}=1$, $k_{21}=1$, $k_{21}=3$.}
  \label{fig.12}
\end{figure} 

By analyzing an ensemble of Gillespie simulations, Figure \ref{fig.12}(a) is obtained. The figure shows the average dynamics and range of the fluctuations. It is worth noting that averaging many fluctuating trajectories from the output of the Gillespie simulations tends to make the average value stationary in the long run. This occurs even when a single realization constantly oscillates. This shows that this approach does not help describe the characteristic dynamics of a single system, as it represents the collective behavior of many systems.
The MFK approach and deterministic Hill function provide the characteristic size of the fluctuations, as shown in Fig.  \ref{fig.12}(b). The characteristic dynamics and size of the fluctuations obtained using the corrected Hill functions are shown in Fig. \ref{fig.12}(c). Again, we see that the reaction rate fluctuations have feedback effects that compensate for fluctuations in other variables. 

In fig. \ref{fig.13}, we compare the fluctuations in the system when using the deterministic Hill function, the Hill function with stochastic corrections, and the analytical result obtained in the steady state. It is observed that the Hill function with stochastic corrections provides a better approximation of the system.

\section{Results and conclusions}\label{section6}
The second-order expansion of the reaction rates around the average dynamics of a stochastic system allowed us to describe the mesoscopic dynamics and fluctuations of the system more accurately than the Linear Noise Approximation and commonly used dissipation fluctuation theorem (\cite{Gar, Scot}). This approximation allowed us to obtain a corrected Hill function that considers the change in the effective reaction rate due to the underlying fluctuations in the ligand-receptor processes. The corrected Hill function can be introduced in a straightforward manner to describe gene regulatory networks, where standard Hill functions are typically used. Because the corrected Hill function depends on the average and variance of the concentrations, it is not introduced in the master equation but in the deterministic description (mesoscopic dynamics) and characterizes the size of its fluctuations. 
 We used the proposed algorithm to analyze the dynamics and fluctuations in gene regulatory networks and compared the different approximations.   
Specifically, by examining the size of the fluctuations using the corrected Hill function and the fluctuations obtained by the Gillespie algorithm and the MFK approach with the traditional Hill function, it was observed that fluctuations in Hill-type propensity rates have a feedback effect that reduces the intrinsic fluctuations in the dynamics of the species involved in gene regulatory networks. In particular, this effect was observed in both the repressilator and the activator-repressor clock.

Using the MFK approach together with the Hill function with stochastic corrections will allow us to study the intrinsic fluctuation effect in more complex and larger network structures with a better approximation, as it is significantly more computationally efficient than the Gillespie algorithm.

\section{Acknowledgments}
\noindent The author Manuel E. Hernández-García acknowledges the financial support of CONAHCYT through the program "Becas Nacionales 2021".\\
Jorge Velázquez-Castro acknowledges financial support of VIEP-BUAP through project 00226-VIEP 2023.

\section*{Declarations}
The authors declare that there is no conflict of interest regarding the publication of this article.

All data generated or analyzed during this study are included in this published article.

\appendix




\section{Calculation of the stationary distribution} \label{A0}
We assumed that the fast reactions reached their stationary state before the slow reactions. Thus, when describing the dynamics of slow reactions, we assume that fast reactions are already stationary. In this section, we describe the methodology used to calculate the stationary distribution of the master equation. 

The stationary distribution is the solution of (\ref{2.4}) when its LHS is set to zero
\begin{align}
    0=& \sum _i \left( t_i^{-}(\mathbf{S}+\Gamma_{i}) P(\mathbf{S}+\Gamma_{i},t) - t_i^{+}(\mathbf{S}) P(\mathbf{S},t) \right. \nonumber \\
    &+\left.t_i^{+}(\mathbf{S}-\Gamma_{i}) P(\mathbf{S}-\Gamma_{i},t) - t_i^{-}(\mathbf{S}) P(\mathbf{S},t) \right).  \label{2.5}
\end{align}
Several proposals have been proposed to solve (\ref{2.5}) \cite{Gar, Kurtz}. We followed a procedure similar to \cite{Kurtz} with a slight variation. As shown in \cite{Kurtz}, to find the stationary distribution we need our system to be at least weakly  reversible and has zero deficiency can be expressed in the form
\begin{align}
    P(\textbf{S})_S=\prod_i P_i(S_i). \label{2.6}
\end{align}
Where $P_{i}(S_{i})$ is the steady probability of finding $S_{i}$ molecules of species $i$. 
Then we proceed as follows,
\begin{enumerate}
    \item [1.-] We define 
            \begin{align}
                \nu^{+}_i(\mathbf{S})&= t_i^{+}(\mathbf{S}) P(\mathbf{S}), &
                \nu^{-}_i(\mathbf{S})&= t_i^{-}(\mathbf{S}) P(\mathbf{S}).
            \end{align}
    \item [2.-] By complex balance \cite{Kurtz}, the stationary distribution obey the following system of equations
    {\tiny
           \begin{align}
              &\sum_j \Gamma_{ij} (\nu^{+}_j(\mathbf{S})\theta(\Gamma_{ij})+ \nu^{-}_j(\mathbf{S}) \theta(-\Gamma_{ij})) = \sum_j \Gamma_{ij} (\nu^{+}_j(\mathbf{S}-\Gamma_{j})\theta(-\Gamma_{ij})+ \nu^{-}_j(\mathbf{S}+\Gamma_{j})\theta(\Gamma_{ij})) \label{2.8} .
           \end{align}}
where $\Gamma_{ij}$ is the stoichiometric matrix, and $\theta(x)$ is the Heaviside step function. 
    \item [3.-] There might be equations that are linearly dependent on others, meaning that depending on the particular form of $\Gamma_{ij}$ we could have more equations than unknowns in the system (\ref{2.8}). In this case, we can transform the similarity to represent $\Gamma_{ij}$ in its reduced-echelon form $\Gamma_{ij}'$. If there are linearly dependent columns, this implies that there are conserved quantities. Thus the linear independent system of equation is
    
    {\tiny
           \begin{align}
              &\sum_j \Gamma_{ij}' (\nu^{+}_j(\mathbf{S})\theta(\Gamma_{ij}')+ \nu^{-}_j(\mathbf{S}) \theta(-\Gamma_{ij}')) = 
               \sum_j \Gamma_{ij}' (\nu^{+}_j(\mathbf{S}-\Gamma_{j}')\theta(-\Gamma_{ij}')+ \nu^{-}_j(\mathbf{S}+\Gamma_{j}')\theta(\Gamma_{ij}')) . \label{2.9}  
           \end{align}} 
    \item [4.-] Substituting (\ref{2.6}) in (\ref{2.9}) and taking the average over all variables except a single specie $S_i$ we obtain an equation for $P_{i}(S_i)$
    
    {\footnotesize
        \begin{align}
              &\sum_j \Gamma_{ij}' (\braket{t^{+}_j(\mathbf{S})}_{l \neq i} P_i(S_i)\theta(\Gamma_{ij}')+ \braket{t^{-}_j(\mathbf{S})}_{l \neq i} P_i(S_i) \theta(-\Gamma_{ij}')) = \nonumber \\
               &\sum_j \Gamma_{ij}' \braket{t^{+}_j(\mathbf{S})}_{l \neq i} P_i(S_i-\Gamma_{ij})\theta(-\Gamma_{ij}')+ \braket{t^{+}_j(\mathbf{S})}_{l \neq i} P_i(S_i+\Gamma_{ij})\theta(\Gamma_{ij}')) . 
         \end{align}}
This procedure can be repeated for the remaining species of the system.
After solving for each species it is possible to find the stationary distribution around a stationary state 
\begin{align}
             P(\mathbf{S})= \prod_i M \frac{c_i^{S_i}}{S_i !} \prod_l \delta _{\sum_{i} \gamma_{li} S_i - N_l,0}.
\end{align}
Here the coefficient $M$ is a normalization constant, $c_i$ characterizes the mean concentration of specie $i$, $\gamma_{li}$ are the null vectors of the stoichiometric matrix, $N_{l,0}$ are the conserved quantities of the system.  We introduce the Kronecker delta to consider that there could be some dependent variables in $\mathbf{S}$ labeled by index $l$.
\end{enumerate}

\section{Stochastic derivation of Hill function} \label{secA1}
Now, our objective is to derive the Hill function from a stochastic process. If we remember that the derivation or deterministic function relies on the steady state, we will also make a similar choice, although in this case, we will look at the stationary distribution. We use the method presented in Appendix \ref{A0}, and the stationary distribution allows us to determine the fluctuations of the system. 

We suppose that we have a ligand $p$ to which $n$ receptors $e$ can bind, forming $s=pe_{n}$, so our system is described by the following reversible processes             
\begin{align}
p + n e \stackbin[k_{-}]{k_{+}}{\rightleftarrows} s.  \label{hes5}
\end{align}
Note that the derivations made in this section are valid only when the previous reactions are found in the system. As we have already explained, the master equation of a chemical network can be obtained from the formalism developed in section 2 (multivariable birth-death processes), for this, we first define our stoichiometric coefficient matrices and the stoichiometric matrix        
\begin{align}
 \alpha_{ij}&= \begin{pmatrix}
		n& 1 & 0
	\end{pmatrix} ,    &
 \beta_{ij}&=  \begin{pmatrix}
		0& 0 & 1
	\end{pmatrix},  &
 \Gamma_{ij}&= \begin{pmatrix}
		-n \\
		 -1 \\
		 1
	\end{pmatrix}, \label{5.8}
\end{align}
With the help of these, we calculate the propensity rates and we get
\begin{align}
    t_1^{+}&= k_{+} p \frac{e!}{(e-n)!} \frac{1}{\Omega^{n+1}}, & t_1^{-}= k_{-} s\frac{1}{\Omega}.     \label{5.10}
\end{align}
(The subscript 1 corresponds to the reaction on the left of (\ref{hes5}), while 2 corresponds to the right.) To find the master equation of the system, we substitute all the quantities we have calculated, we finally get our master equation

{\footnotesize
\begin{align}
    \dot P(e,p,s,t)&= k_{-}(s+1) P(e-n,p-1,s+1,t)- k_{+}p \frac{e!}{(e-n)!} \frac{1}{\Omega^{n}} P(e,p,s,t) \nonumber \\
    &+ k_{+}(p+1) \frac{(e+n)!}{e!} \frac{1}{\Omega^{n}} P(e+n,p+1,s-1,t) - k_{-} s P(e,p,s,t) , \label{hes1}
\end{align}}
If we consider that our system achieves equilibrium rapidly, it is sufficient to examine the stationary distribution, in which the previous relationship becomes zero. To discover the stationary distribution, we employ the method we have previously presented. Since we possess a conserved quantity, the total number of ligands $p$ and remains constant, equal to $N_0= p+s$. For ease of connection with the deterministic aspect, we define $N_0= N_{0d} \Omega$, where $N_{0d }$ represents the deterministic initial condition. By substituting the constant, we obtain

{\footnotesize
\begin{align}
    0&= k_{-}(s+1) P_S(e-n,s+1)- k_{+}(N_0-s) \frac{e!}{(e-n)!} \frac{1}{\Omega^{n}}P_S(e,s) \nonumber \\
    &+ k_{+}(N_0-s+1) \frac{(e+n)!}{e!} \frac{1}{\Omega^{n}}P(e+n,s-1) - k_{-} s P_S(e,s) .
\end{align}}
Continuing with the method,  we will finally have the stationary distribution

{\small
\begin{align}
    P_S(s)&= \frac{1}{  \left(1+  \frac{k_{+}}{k_{-}} \frac{1}{\Omega^n} \langle {\frac{e!}{(e-n)!}} \rangle_e \right)^{N_0} } \left( \frac{k_{+}}{k_{-}} \frac{1}{\Omega^n} \langle {\frac{e!}{(e-n)!}} \rangle_e  \right)^{s}\frac{ N_0! }{s!(N_0 - s)!}. \label{5.18}
\end{align}}
This distribution is a variant of a binomial, whereas it indicates the joining and separation of the proteins and enzymes. An important consideration is that our system may be within a network that contains more interactions; therefore, in this case, the steady-state distribution does not consider molecules $e$. With this last equation, we calculate the average of the variable $s$ and $e$,        
\begin{align}
    \braket{s}&= N_0 \frac{\frac{k_{+}}{k_{-}} \frac{1}{\Omega^n} \langle {\frac{e!}{(e-n)!}} \rangle_e }{1+ \frac{k_{+}}{k_{-}}\frac{1}{\Omega^n} \langle {\frac{e!}{(e-n)!}} \rangle_e }.
\end{align}
If we remember that the Hill function can also be calculated as $H= \frac{\braket{s}}{\braket{s+p}}= \frac{\braket{s}}{N_0}$, supporting us from the two previous relationships we have
{\small
\begin{align}
    H= \frac{ \frac{k_{+}}{k_{-}}\frac{1}{\Omega^n} \langle {\frac{e!}{(e-n)!}} \rangle_e }{1+ \frac{k_{+}}{k_{-}} \frac{1}{\Omega^n} \langle {\frac{e!}{(e-n)!}} \rangle_e } \approx \frac{  \frac{1}{\Omega^{n}} \braket{e}^{n}}{\frac{k_{-}}{k_{+}}+   \frac{1}{\Omega^{n}}\braket{e}^{n}} = \frac{\hat{e}^n}{K^{n}+\hat{e}^n}, \label{Hill1}
\end{align}}
Thus, we recover the deterministic Hill function with $K^n=\frac{k_{-}}{k_{+}}$ and $\hat{e}=\frac{\braket{e}}{\Omega }$. We realize that the deterministic Hill function can also be derived from the stochastic formalism, and we can calculate how our function fluctuates in the stationary state and calculate the fluctuations of $\frac{s} {N_0}$ because this variable is, in fact, the Hill function. Using the steady-state distribution, the fluctuations of the Hill function are as follows:
\begin{align}
    \eta(H)=\frac{1}{\Omega}\sqrt{{\sigma^2\left(\frac{s}{N_0}\right)}}= \frac{1}{\Omega} \sqrt{\frac{H(1-H)  }{\Omega N_0^d K^n  }} , \label{Hill2}
\end{align}
We used $N_0= N_{0d} \Omega$, in which it can be seen that when the system becomes sufficiently large, the fluctuations disappear because $\Omega$ appears in the denominator; when $H$ approaches a value of 0 or 1, the fluctuations become small. In other words, because the concentration $\hat{e}$ is small (the value of $H$ is close to 0), few can bind to the receptor, creating a small fluctuation in $\hat{s}$, whereas when the value of $\hat{e}$ increases (the value of $H$ approaches 1), the system has many receptors but also many ligands available to bind to receptors, so the fluctuations become small. Note that when $H=\frac{1}{2}$ the fluctuations reach their maximum value. 

A similar derivation can be made for the case of a repressor, in this case, we would analyze a set of biochemical reactions similar to those we started in this section, and since the process is very similar (we omit the calculations) the Hill function would be obtained for a repressor and the same magnitude of the fluctuations although these would be given by
\begin{align}
    \eta(D)=\frac{1}{\Omega}\sqrt{{\sigma^2\left(\frac{s}{N_0}\right)}}= \frac{1}{\Omega} \sqrt{\frac{D(1-D)  }{\Omega N_0^d K^n  }} . 
\end{align}
Because $H+D=1$, we can say that the fluctuations in $H$ and $D$ have the same magnitude. 

\section{Hill function in a stochastic process} \label{C1}
A methodology that explains how to properly introduce a Hill function in any stochastic biochemical reaction, such that the result can be generalized to other types of chemical reaction networks, is needed. To carry out the derivation, we rely on the ideas already presented up to this point and \cite{Uri}, where a separation is made between slow and fast variables. For our derivation, we first suppose that the probability distribution of our system is composed of the multiplication of a stationary part and a dynamic part, in such a way that we will have something of the following form
\begin{align}
    P(\mathbf{x},t)= P_s(x_1,..,x_s)P(x_{s+1},...,x_l,t),
\end{align}
in \cite{Uri} conditional probabilities are used, but because we directly consider that a part of the process is stationary, a separation between them can be made; in this case, we can assume that it is stationary because the concentrations of the variables associated with it tend to be very fast and asymptotically to the stationary value (the behavior of the Hill function). Under these considerations, the master equation of the system will be divided into a dynamic and stationary part  \cite{Uri, Gout, Kim2, Cao} that will be as follows

{\tiny
\begin{align}
   P_s(S_1,..,S_s) \frac{dP(S_{s+1},...,S_j,t)}{dt}=&\Omega \sum _i \left( t_i^{-}(S_j+\Gamma_{ji}) P_s(S_1+\Gamma_{ii},..,S_s  +\Gamma_{si}) P(S_{s+1}+\Gamma_{s+1,i},... S_{j}+\Gamma_{j,i},t) \right. \nonumber \\
   &- t_i^{+}(S_j) P_s(S_1,..,S_s) P(S_{s+1},...,S_j,t)  \nonumber \\
    &+t_i^{+}(S_j-\Gamma_{ji}) P_s(S_1-\Gamma_{ii},..,S_s  -\Gamma_{si}) P(S_{s+1}-\Gamma_{s+1,i},... S_{j} \nonumber \\
    &-\left. \Gamma_{j,i},t) - t_i^{-}(S_j) P_s(S_1,..,S_s) P(S_{s+1},...,S_j,t) \right), \nonumber \\
    P(S_{s+1},...,S_j,t) \frac{dP_s(S_1,..,S_s)}{dt}  =& \Omega \sum _{i=rap} \left( t_i^{-}(S_j+\Gamma_{ji}) P_s(S_1+\Gamma_{ii},..,S_s  +\Gamma_{si}) P(S_{s+1}+\Gamma_{s+1,i},... S_{j}+\Gamma_{j,i},t) \right. \nonumber \\
    &- t_i^{+}(S_j) P_s(S_1,..,S_s) P(S_{s+1},...,S_j,t)  \nonumber \\
    &+t_i^{+}(S_j-\Gamma_{ji}) P_s(S_1-\Gamma_{ii},..,S_s  -\Gamma_{si}) P(S_{s+1}-\Gamma_{s+1,i},... S_{j} \nonumber \\
    &-\left.\Gamma_{j,i},t) - t_i^{-}(S_j) P_s(S_1,..,S_s) P(S_{s+1},...,S_j,t) \right)=0. \label{4.50}
\end{align}}
(Note that only fast reactions are considered in the second equation) Remember that the stationary part is the one associated with the Hill function, that is, this part is given by the following reaction
\begin{align}
p + n e \stackbin[k_{-}]{k_{+}}{\rightleftarrows} pe_n. 
\end{align}
These reactions are generally fast; therefore, the Hill function is determined if a deterministic analysis is performed. However, because we are using a stochastic approach, we would have to proceed in another way. Assuming that we are already in the stationary state, from these reactions and with the help of (\ref{4.50}) we would obtain something similar to the following

{\footnotesize
\begin{align}
    k_{-}(s+1) P_s(e-n,p-1,s+1)- k_{+}p \frac{e!}{(e-n)!} \frac{1}{\Omega^{n}} P_s(e,p,s)  + \nonumber \\
     k_{+}(p+1) \frac{(e+n)!}{e!}\frac{1}{\Omega^{n}} P_s(e+n,p+1,s-1) - k_{-} s P_s(e,p,s)=0. \label{hes4}
\end{align}}
($e$, $p$ and $s$ are the number of molecules $e$, $p$ and $pe_n$ respectively) Now, the question is what to do with this equation and how the Hill function would emerge, there are two methods, as we will see below.

\subsection{Exact Hill function}
In this first derivation, from the equation (\ref{hes4}) a stationary distribution is obtained, we use the method that we explained to obtain it,

{\small
\begin{align}
    P_s(p,s)&= \left(\frac{1}{\Omega^{n}} \frac{k_{+}}{k_{-}}  \langle {\frac{e!}{(e-n)!}} \rangle_e \right)^{s}\frac{N_0 ! }{s! (N_0 -s )!} P_s(0).
\end{align}}
(For more details on obtaining this expression, please refer to the Appendix \ref{secA1}.) Now we average the first equation of (\ref{4.50}) with respect to this stationary distribution, we define the average with respect to the stationary part as $\braket{{t}^{\pm} }_S$ of this way our first equation becomes

{\tiny
\begin{align}
    \frac{dP(S_{s+1},...,S_j,t)}{dt}= \Omega \sum _i &\left( \braket{t_i^{-}(\mathbf{S}+\Gamma_{i})}_s  P(S_{s+1}+\Gamma_{s+1,i},... S_{j}+\Gamma_{j,i},t) - \braket{t_i^{+}(\mathbf{S})}_s  P(S_{s+1},...,S_j,t) \right. \nonumber \\
    &+\left.\braket{t_i^{+}(\mathbf{S}-\Gamma_{i})}_s  P(S_{s+1}-\Gamma_{s+1,i},... S_{j}-\Gamma_{j,i},t) - \braket{t_i^{-}\mathbf{S}}_s  P(S_{s+1},...,S_j,t) \right), 
\end{align}}
This equation was used to model the dynamics of this type of system; only now are the reaction rates averaged with respect to the stationary part, for more details on the developments that can be reviewed \cite{Uri}. Hill functions are expected to appear in the averages with respect to the stationary part, as observed in the case of the Toggle Switch. However, because some averages would appear within this expression, it is not recommended to use Gillespie's algorithm; however, we can still find the average of the concentrations and quantify the intrinsic fluctuations if the MFK approach is used.

\subsection{Hill function variants}

\begin{table}[h!]
\centering
\begin{tabular}{ |p{3.5cm}|p{3cm}|p{3cm}|  }
\hline
\multicolumn{3}{|c|}{Assumptions} \\
\hline
System Size & Distribution & Hill Function Type \\
\hline
X & X & Stochastic \\
X & \checkmark   & Semi-stochastic \\
\checkmark & X & Semi-deterministic \\
\checkmark & \checkmark & Deterministic \\
\hline
\end{tabular} 
\caption{Assumptions table.}  \label{tabla1}
\end{table}

In the case in which we want to do Gillespie-type simulations, and also explain how the deterministic Hil function appears, we can derive another type of Hill function, from (\ref{hes4}) we find the following two relations

{\tiny
\begin{align}
   s+1=&p \frac{1}{K^n} \frac{e!}{(e-n)!} \frac{1}{\Omega^{n}} \frac{P(e,p,s)}{P(e-n,p-1,s+1)},  &
    s= (p+1) \frac{1}{K^n} \frac{(e+n)!}{e!} \frac{1}{\Omega^{n}} \frac{P(e+n,p+1,s-1)}{P(e,p,s)}, \label{hes2}
\end{align}}
We define $\frac{k_{-}}{k_{-}}= K^n $. Several assumptions can be made to determine the Hill function, including the size of the system and/or shape of the distribution (these conditions were chosen to treat the previous equations as algebraic), as outlined in Table \ref{tabla1}. In this table, we have four types of Hill functions; its type depends on the type of assumptions that are considered: in the case that we do not make any, we have a stochastic case; when considerations are made in the probability distribution, we have a semi-stochastic case, and so on. These names were chosen because of the degree of relevance of the components to each formalism. For example, in the stochastic case, one mainly considers the master equation, which is a differential equation of distributions. \\

\textbf{Semi-deterministic case} \\

First, we analyze the semi-deterministic case, assuming that the system is sufficiently large. The probability distribution can then be approximated as $\frac{P(e,p,s)}{P(e-n,p-1, s+1)} \approx 1$. We consider two cases because we have two expressions in (\ref{hes2}) and $p+s=N_0$:
\begin{enumerate}
    \item[a)]For this case, we use the first equation of (\ref{hes2}), $s$ and $N_0$ are very large (because the system is large) so $s+1 \approx s $ and also $\frac{P(e,p,s)}{P(e-n,p-1,s+1)} \approx 1$, so we would have
    \begin{equation}
        s=p \frac{1}{K^n} \frac{e!}{(e-n)!} \frac{1}{\Omega^{n}},
    \end{equation}
   substituting in the definition of the Hill function $H= \frac{s}{s+p}$ we obtain,
    \begin{equation}
        H_1=\frac{e!}{\Omega^n K^n (e-n)!+ e!} \label{hes3}.
    \end{equation}
    \item[b)] For this other case, we use the second equation of (\ref{hes2}), $p$ and $N_0$ are very large (because the system is large) so $p+1 \approx p $, said of otherwise $s$ has to be very small because we have the relation $N_0=s+p$, furthermore, we suppose $\frac{P(e+n,p+1,s-1)}{P(e, p,s)} \approx 1$, so we would have
    \begin{equation}
        s=p \frac{1}{K^n} \frac{(e+n)!}{e!}\frac{1}{\Omega^{n}},
    \end{equation}
    Substituting in the definition of the Hill function we get,
    \begin{equation}
        H_2=\frac{(e+n)!}{\Omega^n K^n e!+ (e+n)!} \label{he4}.
    \end{equation}
\end{enumerate}
In this way we have obtained our Hill function for two regions, so we can say that one is valid when $s$ is small and the other when it is large, so we define a Hill function for all possible values of $s$ as follows

{\footnotesize
\begin{align}
    H_{sd}= \frac{e!}{\Omega^n K^n (e-n)!+ e!} +\left( \frac{(e+n)!}{\Omega^n K^n e!+ (e+n)!}-\frac{e!}{\Omega^n K^n (e-n)!+ e!} \right) \theta\left(\frac{1}{2}-\frac{s}{N_0}\right),
\end{align}}
A Heavennside function was introduced to separate the two regions of the function, which relies on $H_1$ or $H_2$ depending on the value of $s$, thus covering all possible values of $s$. We obtained our semi-deterministic Hill function, which is semi-deterministic because the probability considerations are almost deterministic. Also, when $e$ and/or $\Omega$ get very large    
\begin{equation}
    H_{sd}\approx H_d= \frac{\frac{e^n}{\Omega^n}}{K^n + \frac{e^n}{\Omega^n}},
\end{equation}
A Heavennside function was introduced to separate the two regions of the function, which relies on $H_1$ or $H_2$ depending on the value of $s$, thus covering all possible values of $s$. We obtained our semi-deterministic Hill function, which is semi-deterministic because the probability considerations are almost deterministic. In addition, when $e$ and/or $\Omega$ become very large.    \\

\textbf{Stochastic case} \\

In the case that we want to use a stochastic Hill function, we can construct a completely stochastic one, for this we return to the relations found in (\ref{hes4}) but without making any assumptions, we operate in a similar way to the case above, to cover the different values of $s$, we finally obtain our stochastic Hill function

{\tiny
\begin{align}
    H_s&= \frac{e!}{\Omega^n K^n (e-n)!\frac{P(e-n,s+1)}{P(e,s)}+ e! } \nonumber  \\ 
    &+ \left( \frac{(e+n)!}{\Omega^n K^n e! \frac{P(e,s)}{P(e+n,s-1)}     + (e+n)!}-\frac{e!}{\Omega^n K^n (e-n)!\frac{P(e-n,s+1)}{P(e,s)}+ e! } \right) \theta\left(\frac{1}{2}-\frac{s}{N_0}\right).
\end{align}}
This is the stochastic Hill function, which depends on the probability distribution of the system and, using this function as defined within the master equation, becomes more complex.\\

\textbf{Semi-Stochastic case} \\

Because it is more complicated to use the Hill function in a stochastic process, we approximated the probabilities. So we suppose the following
\begin{align}
    \frac{P(e+e_1,s+s_1)}{P(e,s)} \approx e^{-\left( \frac{e_1+ s_1}{\Omega} \right)},
\end{align}
We used this approximation because when the system is large, the binomial distribution that appears in the steady state tends to be a Poisson distribution, and when $\Omega$ is sufficiently large, we find the deterministic assumption of cases a) and b). Substituting these relations in the previous equation we obtain our semi-stochastic Hill function, because we make a consideration on the distribution $P(e,s)$,

{\tiny
\begin{align}
    H_{ss}&= \frac{e!}{e^{\left( \frac{n-1}{\Omega} \right)}\Omega^n K^n (e-n)!+ e! }  + \left( \frac{(e+n)!}{e^{\left( \frac{n-1}{\Omega} \right)} \Omega^n K^n e!  + (e+n)!}-\frac{e!}{e^{\left( \frac{n-1}{\Omega} \right)} \Omega^n K^n (e-n)!+ e! } \right) \theta\left(\frac{1}{2}-\frac{s}{N_0}\right).
\end{align}}

\begin{figure} [h!t]
  \begin{subfigure}{\linewidth}
   \includegraphics[width=.24\textwidth]{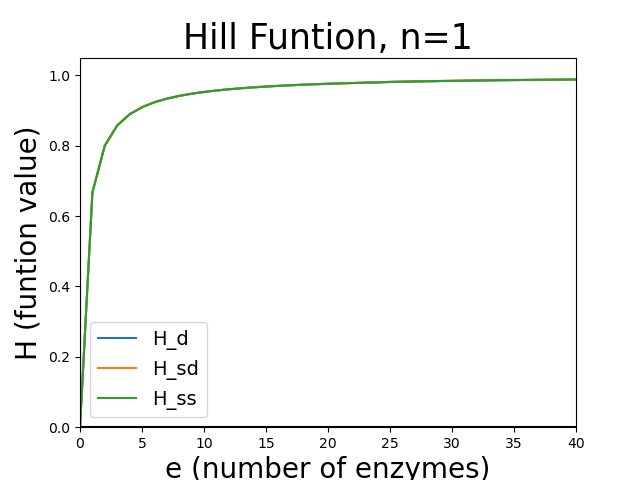}\hfill
\includegraphics[width=.24\textwidth]{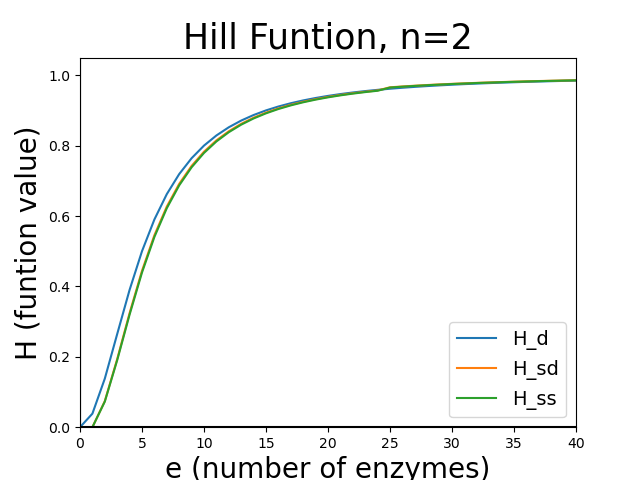}\hfill
\includegraphics[width=.24\textwidth]{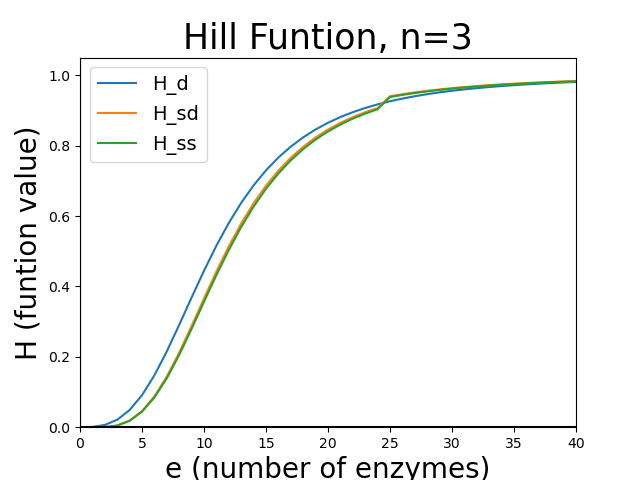}\hfill
\includegraphics[width=.24\textwidth]{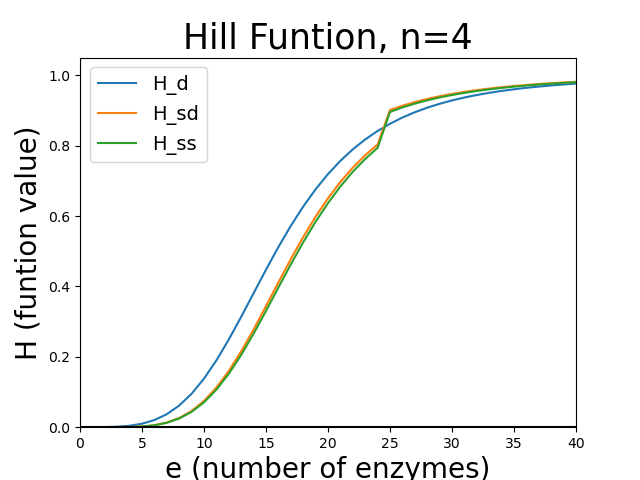}
\end{subfigure}
\caption{Hill functions. In this figure the deterministic $H_d$, semi-deterministic $H_{sd}$ and semi-stochastic Hill function $H_{ss}$ is plotted for different values of $n$, with $K=0.01$, $\Omega=50$.}
  \label{fig28}
\end{figure}

We call this function the semi-stochastic Hill function because we made a consideration only on the probability distribution, this function coincides with the semi-deterministic when $n=1$ or when the value of $\Omega$ is very great,
\begin{align}
    \lim_{\Omega \to \infty} H_{ss} = \frac{\frac{e^n}{\Omega^n}}{K^n + \frac{e^n}{\Omega^n}}=H_d,
\end{align}
Therefore, to perform simulations of a certain stochastic system, and when the system size is small, we recommend using the semi-stochastic Hill function because it represents the system more faithfully. This function is considerably easier to use in the stochastic simulations. 

To observe the behavior of the Hill functions that we calculated, $H_d$, $H_{sd}$ and $H_{ss}$, we made some graphs that can be seen in figure \ref{fig28}, in which we notice that when $n=1$ it is observed that the three functions are practically the same. When the value of $e$ increases, the values of the three are almost identical, whereas at small values, there is a large difference, as shown in the figures. Another feature that they share is that they all practically become step functions when $n$ is large. However, the location at which the step varies, because factorials appear is semi-deterministic or semi-stochastic. 

The considerations we have made regarding the appearance of Hill functions may not be completely fulfilled (even more so when the size of the system is small). Therefore, if one wants a process that is as realistic as possible, it is better to use the relationships presented at the beginning of this subsection. However, other complications can occur if the constants are unknown. Therefore, the best approach is to use an exact derivation.

\section{General Hill function} \label{B1}

\begin{figure} [h!t]
  \centering
\includegraphics[width=1\textwidth]{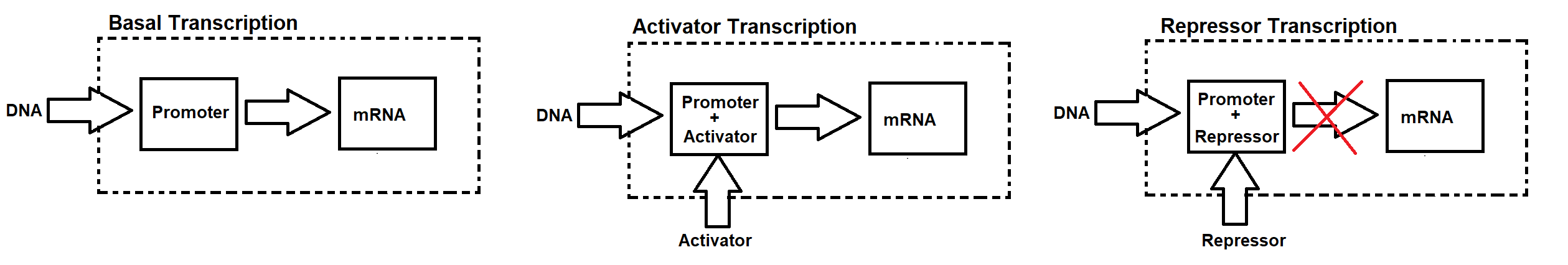}
  \caption{ Genetic transcription.} \label{a1}
\end{figure} 

In this appendix, we derive the Hill function exclusively when several proteins act as transcription factors. Figure \ref{a1} shows that there are three ways in which mRA is synthesized: basal transcription is always present, whereas when transcription factors appear, transcription is faster (activator) or inhibited (repressor). This model was inspired by \cite{VecchioM}. For this, we assume that two types of proteins $P_1$ and $P_2$ act as transcription factors as activators and suppressors respectively, all the processes involved are given in the following biochemical reactions

{\small
\begin{enumerate}
    \item [] Basal union: $G + P_r \stackbin[\lambda_1^{-}]{\lambda_1^{+}}{\rightleftarrows}  GP_r$.

    \item [] Activator transcription factor: $n_1P_1 + P_r \stackbin[k_{1}^{-}]{k_{1}^{+}}{\rightleftarrows}  P_{rp_1}$.
    
    \item [] Suppressor transcription factor: $n_2 P_2 + P_r \stackbin[k_{2}^{-}]{k_{2}^{+}}{\rightleftarrows}  P_{rp_2}$.
    
    \item [] Binding with transcription factors: $G + P_{rp_1} \stackbin[\lambda_2^{-}]{\lambda_2^{+}}{\rightleftarrows}  GP_{rp_1}$.
    
    \item [] Basal mRNA synthesis: $GP_r \stackbin{\delta_{1}}{\rightarrow}  R$.
    
     \item [] mRNA synthesis with transcription factors: $GP_{rp_1} \stackbin{\delta_{2}}{\rightarrow}  R$.
\end{enumerate}}

$G$ represents the gene, $P_r$ is the promoter, $R$ is the mAR, $GP_r$ is the gene binding to the promoter, $P_{rp_1}$ and $P_{rp_2}$ indicate when the transcription factor has joined the promoter. The first reaction at the top is a reversible process, indicating that the gene binds to the promoter regardless of the presence of transcription factors. The second reaction is also a reversible process in which transcription factors bind to the promoter in such a way that they create a new molecule that accelerates mRNA synthesis. In the third reaction, a transcription factor creates a new molecule that cannot bind to a gene. In the fourth reaction, $G$ binds to the $P_{rp_1}$ molecule. The last two reactions indicate how mRNA is synthesized by a one-way process. 

We used the law of mass action to build a set of differential equations that describe the concentrations of our system, where $g$ represents the gene concentration, $p_r$ is the promoter concentration, $r$ is the mAR concentration, $ s_1$ is the concentration of $GP_r$, $s_2$ and $s_3$ at concentrations of $P_{rp_1}$ and $P_{rp_2}$ respectively, $s_4$ is the concentration of $GP_{rp_1}$. The differential equations we obtain are the following,

{\tiny
\begin{align}
    \frac{d s_1}{dt}=& \lambda_1^{+} gp_r - \lambda_1^{-}s_1, &
    \frac{d s_2}{dt}=& k_1^{+}  p_1^{n_1}p_r - k_1^{-}s_2 - \lambda_2^{+} gs_2 + \lambda_1^{-}s_4, \nonumber \\
    \frac{d s_3}{dt}=& k_2^{+} p_2^{n_2} p_r - k_2^{-}s_3, &
    \frac{d s_4}{dt}=&  \lambda_2^{+} gs_2 - \lambda_1^{-}s_4, \nonumber \\
    \frac{d p_r}{dt}=& -\lambda_1^{+} gp_r + \lambda_1^{-} - k_1^{+} p_1^{n_1} + k_1^{-}p_r - k_2^{+} p_2^{n_2} + k_2^{-}p_r, &
    \frac{d r}{dt}=& \delta_1 s_1 + \delta_2 s_4, \label{hg0}
\end{align}}
From these differential equations, we can see that we have a conserved quantity and that the initial concentration of the promoters is constant, $p_0= p_r + s_1+s_2+s_3 + s_4$. Of the reactions that we have, we consider that the first 4 are practically in equilibrium, so the first four differential equations are considered in equilibrium, from which we obtain the following conditions
\begin{align}
    s_1=& \frac{\lambda_1^{+}}{\lambda_1^{-}} g p_r= \lambda_1 g  p_r , &
    s_2=& \frac{k_1^{+}}{k_1^{-}} p_1^{n_1} p_r = \frac{p_1^{n_1}}{ K_1^{n_1}} p_r, \nonumber \\
    s_3=& \frac{k_2^{+}}{k_2^{-}} p_2^{n_2} p_r = \frac{p_2^{n_2}}{ K_2^{n_2}}p_r, &
    s_4=& \frac{\lambda_2^{+}}{\lambda_2^{-}} g s_2 = \lambda_2 g  s_2. \label{hg1}
\end{align}
Our goal is to find a relationship that describes $s_1$ and $s_4$ in terms of transcription factors and basal synthesis since these are closely related to mRNA synthesis as can be seen in (\ref{hg0} ), then with the help of the above relations we can write the following
\begin{align}
    H_1=& \frac{s_1}{p_0}= \frac{s_1}{p_r+s_1+s_2+s_3+s_4}, &
    H_2= \frac{s_4}{p_0}= \frac{s_4}{p_r+s_1+s_2+s_3+s_4},
\end{align}
when evaluating the quantities obtained in (\ref{hg1}) these relations we obtain 

{\footnotesize
\begin{align}
    H_1=& \frac{\lambda_1 g}{1 + \lambda_1 g + \frac{p_1^{n_1}}{ K_1^{n_1}} +  \frac{p_2^{n_2}}{ K_2^{n_2}} + \lambda_2 g  \frac{p_1^{n_1}}{ K_1^{n_1}} }, &
    H_2=  \frac{\lambda_2 g  \frac{p_1^{n_1}}{ K_1^{n_1}}}{1 + \lambda_1 g + \frac{p_1^{n_1}}{ K_1^{n_1}} +  \frac{p_2^{n_2}}{ K_2^{n_2}} + \lambda_2 g  \frac{p_1^{n_1}}{ K_1^{n_1}} }.
\end{align}}
Remember that the synthesis of the mRNA is given by the term $\delta_1 s_1 + \delta_2 s_4$, so now we will have
\begin{align}
    \delta_1 s_1 + \delta_2 s_4 = \delta_1 p_0 \frac{s_1}{p_0} + \delta_2  p_0 \frac{s_4}{p_0} =  \delta_1 p_0 H_1 + \delta_2  p_0 H_2 = H,
\end{align}
where
\begin{align}
    H= \frac{ \delta_1 p_0 \lambda_1 g + \delta_2  p_0 \lambda_2 g  \frac{p_1^{n_1}}{K_1^{n_1}}}{1 + \lambda_1 g + \frac{p_1^{n_1}}{ K_1^{n_1}} +  \frac{p_2^{n_2}}{ K_2^{n_2}} + \lambda_2 g  \frac{p_1^{n_1}}{ K_1^{n_1}} } = \frac{ \beta_0 K_0 + \beta_{11}  \frac{p_1^{n_1}}{K_1^{n_1}}}{1 + K_0 + \beta_{21}\frac{p_1^{n_1}}{ K_1^{n_1}} + \beta_{22} \frac{p_2^{n_2}}{ K_2^{n_2}} } , 
\end{align}
($\beta_0 = \delta_1 p_0 $, $K_0 = \lambda_1 g $, $\beta_{11}= \delta_2 p_0 \lambda_2 g $, $\beta_{21}=(1 + \lambda_2 g )$, $ \beta_{22}=1$). The function $H$ that we defined corresponds to the Hill function of our system, which indicates how the ARm is synthesized in terms of transcription factors $P_1$ and $P_2$, in addition to a basal constant $K_0$. $\beta_0 K_0$ denotes the basal synthesis rate. Now we can give a generalization of this expression, for this we suppose that we have a system with $m$ proteins that acts both as activators or suppressors, then our generalized Hill function ($H_g$) will be of the form
\begin{align}
    H_g= \frac{ \beta_0 K_0 + \sum_j  \beta_{1j}  \left(\frac{p_j^{n_j}}{K_j^{n_j}}\right)^q}{1 + K_0 + \sum_j \beta_{2j}  \frac{p_j^{n_j}}{K_j^{n_j}}},
\end{align}
we added $q$ to the numerator because if the protein acts as an activator, $q=1$, whereas if it acts as a suppressor, $q=0$. In this way, we have made a generalization of the Hill function, which will be very useful for analyzing any transcription-translation module in which many proteins participate. We can even introduce stochastic corrections to this Hill function, it will only be necessary to change to $p_j^{n_j}$ as follows
\begin{align}
    p_j^{n_j} \rightarrow p_j^{n_j} + \frac{1}{2} \sigma^2(p_j,p_j) n_j(n_j-1)p_j^{n_j-2},
\end{align}
This change is made according to the manner in which the Hill function with stochastic corrections is obtained.

\end{document}